\documentclass[aps,prd,onecolumn,groupedaddress,nofootinbib,superscriptaddress]{revtex4}  
\usepackage{graphicx,mathtools,tabu}
\usepackage{epstopdf}
\usepackage{amsmath}
\usepackage{amsfonts}
\usepackage[colorlinks=true,citecolor=red,linkcolor=blue,breaklinks=true]{hyperref}
\usepackage{accents}
\usepackage[figure, table]{hypcap}
\usepackage{amssymb}
\usepackage{appendix}
\usepackage{comment}
\usepackage{bbold}
\usepackage{color}
\usepackage{slashed}
\usepackage{subfigure}
\usepackage{setspace}
\usepackage{footnote}
\usepackage{multirow}
\usepackage{braket}
\usepackage[normalem]{ulem}
\usepackage[utf8]{inputenc}
\usepackage{mathrsfs}
\usepackage{longtable}
\usepackage{enumitem}
\usepackage{multirow}
\usepackage{placeins}

\LTcapwidth=\textwidth

\usepackage{url}
\usepackage{wrapfig}
\usepackage{braket}

 \def\be   {\begin{equation}}  
 \def\ee   {\end{equation}}
 \def\ba   {\begin{array}}     
  \def\ea   {\end{array}}
 \def\bea  {\begin{eqnarray}}  
  \def\eea  {\end{eqnarray}}
 \def\bean {\begin{eqnarray*}}  
 \def\eean {\end{eqnarray*}}

  \def\be {\beta}

\begin{document}

\title{On the Effects of Quantum Decoherence in a Future Supernova Neutrino Detection}

\author{Marcos V. dos Santos}
\email[Electronic address: \href{mailto:mvsantos@ifi.unicamp.br}{mvsantos@ifi.unicamp.br} (corresponding author)]{}
\affiliation{Universidade Estadual de Campinas, Instituto de Física Gleb Wataghin, R. Sérgio Buarque de Holanda, 777, Brazil}

\author{Pedro C. de Holanda}
\email{holanda@ifi.unicamp.br}
\affiliation{Universidade Estadual de Campinas, Instituto de Física Gleb Wataghin, R. Sérgio Buarque de Holanda, 777, Brazil}

\author{Pedro Dedin Neto}
\email{dedin@ifi.unicamp.br}
\affiliation{Universidade Estadual de Campinas, Instituto de Física Gleb Wataghin, R. Sérgio Buarque de Holanda, 777, Brazil}
\affiliation{Niels Bohr Institute, University of Copenhagen, DK-2100 Copenhagen, Denmark}

\author{Ernesto Kemp}
\email{kemp@ifi.unicamp.br}
\affiliation{Universidade Estadual de Campinas, Instituto de Física Gleb Wataghin, R. Sérgio Buarque de Holanda, 777, Brazil}
\affiliation{Gran Sasso Science Institute - GSSI, L'Aquila (IT)}

\begin{abstract}

Quantum decoherence effects in neutrinos, described by the open quantum systems formalism, serve as a gateway to explore potential new physics, including quantum gravity. Previous research extensively investigated these effects across various neutrino sources, imposing stringent constraints on spontaneous loss of coherence. In this study, we demonstrate that even within the Supernovae environment, where neutrinos are released as incoherent states, quantum decoherence could influence the flavor equipartition of $3\nu$ mixing. Additionally, we examine the potential energy dependence of quantum decoherence parameters ($\Gamma = \Gamma_0 (E/E_0)^n$) with different power laws ($n = 0, 2, 5/2$). Our findings indicate that future-generation detectors (DUNE, Hyper-K, and JUNO) can significantly constrain quantum decoherence effects under different scenarios. For a Supernova located 10 kpc away from Earth, DUNE could potentially establish $3\sigma$ bounds of $\Gamma \leq 6.2 \times 10^{-14}$~eV in the normal mass hierarchy (NH) scenario, while Hyper-K could impose a $2\sigma$ limit of $\Gamma \leq 3.6 \times 10^{-14}$~eV for the inverted mass hierarchy (IH) scenario with $n=0$~—~assuming no energy exchange between the neutrino subsystem and non-standard environment ($[H,V_p] = 0$). These limits become even more restrictive for a closer Supernova. When we relax the assumption of energy exchange ($[H,V_p] \neq 0$), for a 10~kpc SN, DUNE can establish a $3\sigma$ limit of $\Gamma_8 \leq 4.2 \times 10^{-28}$~eV for NH, while Hyper-K could constrain $\Gamma_8 \leq 1.3 \times 10^{-27}$~eV for IH ($n=0$) with $2\sigma$, representing the most stringent bounds reported to date. Furthermore, we examine the impact of neutrino loss during propagation for future Supernova detection.

\end{abstract}

\maketitle

{
  \hypersetup{hidelinks}
  \tableofcontents
}

\section{Introduction}
\label{sec:intro}

Since the supernova SN1987A, the expectation for the next supernova (SN) neutrino detection has stimulated a number of works proposing tests on new physics in our galaxy, making this event a promising natural laboratory for neutrino physics.

Once $\sim 1$ galactic SN is expected per century \cite{rozwadowska2021rate}, the next event holds the opportunity to break through many aspects of neutrino physics, with capabilities of next-generation detectors, such as DUNE \cite{Acciarri:2016crz, Abi:2018dnh, abi2021supernova}, Hyper-Kamiokande (HK) \cite{Abe:2018uyc} and JUNO \cite{an2016neutrino}, leading to a sensible future measurement, increasing the number of neutrino events from the current few dozen to tens of thousands or more in a SN explosion 10 kpc away from Earth. A typical Core-Collapse SN (CCSN) undergoes three main emission phases to be known (see \cite{mirizzi2016supernova} for a review): neutronization burst, where a high amount of $\nu_e$ is emitted given a rate of $e^-$ capture in the first $\sim 30$ ms after core bounce; accretion, where progenitor mass infall and a high luminosity are expected during roughly $\sim 1$ s; and cooling, a thermal phase where a proto-neutron star cools down via neutrino emission, with $\sim 10$~s of duration.

With the possible future sensitivity and increasing sophistication in SN neutrino simulations \cite{tamborra2017flavor, muller2017supernova, Garching}, 
a precise description of standard neutrino evolution until Earth is been pursued. However, in a SN environment, collective oscillations led by $\nu-\nu$ interactions are a source of high uncertainties, since a definitive solution for the $\nu$ equation of motion has not been achieved, even with many ongoing developments in the topic~\cite{tamborra2021new}. One critical remark is that for the three mentioned SN emission phases, collective oscillations are expected to play an important role only in accretion and cooling, with no significant impact on neutronization burst, given the large excess of $\nu_e$ over other flavors, turning it in a promising environment to test new physics.

In SN neutrino-mixing, if we disregard collective effects, with the only relevant neutrino interaction  being the MSW matter effect, the neutrino flux that comes out of the SN can be treated as an incoherent sum of mass states, and no oscillation is expected\footnote{Given the indistinguishability of $\nu_\mu$ and $\nu_\tau$ ($\bar{\nu}_\mu$ and $\bar{\nu}_\tau$) in the detection, they are generally classified as $\nu_x$ ($\bar{\nu}_x$) in the literature.}. Since $\nu_\alpha$ is generated as a mass state in matter $\nu_{i}^m$, it leaves the SN as a mass state in vacuum $\nu_i$ (for an adiabatic conversion in the SN) until reaching Earth. Despite this expected incoherence, neutrinos coming from a SN could be affected by \textit{quantum decoherence}. In this work, we show the impact of quantum decoherence, or the neutrino evolution from pure to mixed states given a coupling to the environment, in the case of a future SN neutrino detection. 

There are different possible sources of decoherence in neutrino evolution, such as wave packet decoherence, that comes from different group velocities of neutrino mass states disentangling the respective wave packets \cite{kersten2016decoherence, akhmedov2017collective, akhmedov2022damping}, or even Gaussian averaged neutrino oscillation given by uncertainty in energy and path length \cite{ohlsson2001equivalence}. The underlying physics in this work is of a different type and refers to effects induced by propagation in a non-standard environment generated by beyond Standard Model physics, and the term {\it decoherence} used in this work refers to the latter.

The idea of inducing pure elementary quantum states into mixed ones was originally established by Hawking \cite{hawking1975particle} and Bekenstein \cite{bekenstein1975statistical} and discussed by a number of subsequent works \cite{hawking1982unpredictability, ellis1984search, wald1994quantum, unruh1995evolution, ellis1999microscopic}, being attributed to quantum (stochastic) fluctuations of space-time background given quantum gravity effects.
Many authors have given a physical interpretation on the impact of such stochastic quantum gravitational background in neutrino oscillations \cite{lisi2000probing, alexandre2008neutrino, mavromatos2009cpt, dvornikov2019neutrino, stuttard2020neutrino, luciano2021gravitational, stuttard2021neutrino,hellmann2022quantum, hellmann2022searching, ettefaghi2022gravitational}, with expected decoherence being well described by open quantum systems formalism through GKSL (Gorini–Kossakowski–Sudarshan–Lindblad) master equation. 
In particular, in \cite{stuttard2020neutrino}, the authors provided a simple and interesting interpretation of physical scenarios for specific forms of GKSL equation, then we use a similar terminology along this manuscript to guide our choices in the analysis.

Phenomenological studies designed to impose bounds on neutrino coupling to the environment through open quantum systems formalism were investigated in atmospheric \cite{lisi2000probing, coloma2018decoherence}, accelerator \cite{farzan2008reconciling, oliveira2014quantum, oliveira2016dissipative, coelho2017nonmaximal, coelho2017decoherence, carrasco2019probing, carpio2019testing, gomes2019quantum, deromeri2023neutrino}, reactor \cite{gomes2017parameter,deromeri2023neutrino}, and solar \cite{coloma2018decoherence, de2020solar, farzan2023decoherence} neutrinos with different approaches. Only upper limits over quantum decoherence parameters were obtained up to now.

This manuscript is structured as follows: in Section~\ref{sec:qd-in-sn} we show the quantum decoherence formalism, introducing the models to be investigated. In Section~\ref{sec:metho-sim} we discuss the methods to factorize the neutrino evolution and how to use them to impose bounds on quantum decoherence with a future SN detection. We also discuss the role of Earth matter effects. Our results are presented in Section~\ref{sec:future-limits} and in Section~\ref{sec:nu-hierar-measur} we discuss how quantum decoherence could affect the neutrino mass ordering determination. Finally, in Section~\ref{sec:conclusions} we present our conclusions.

\section{Quantum decoherence effects in supernova neutrinos}\label{sec:qd-in-sn}

In this section, we devote ourselves to revisiting quantum decoherence formalism in neutrino mixing and show the impacts on the (already) incoherent SN neutrino fluxes.

\subsection{Formalism}\label{sec:formalism}

Considering the effects of quantum decoherence, we can write the GKSL equation in propagation (mass) basis in vacuum \cite{gorini1976completely, lindblad1976generators}

\begin{equation}\label{eq:gksl}
    \frac{d\rho}{dt} = -i[H,\rho] + \mathcal{D}(\rho)
\end{equation}
where $\mathcal{D}(\rho) = \sum_p^{N^2-1} (V_{p} \rho V_{p}^\dagger - \frac{1}{2} \{V_{p}^\dagger V_{p}, \rho\})$ is a dissipation term, representing the neutrino subsystem coupling to the environment. If (\ref{eq:gksl}) is a general equation of motion to describe $\nu$ propagation and a non-standard effect induces a non-null $\mathcal{D}(\rho)$, we require an increase of von Neumann entropy in the process, which can be achieved imposing $V_p = V_p^\dag$ \cite{benatti1988entropy}. It is also possible to write the dissipation term at the r.h.s. of (\ref{eq:gksl}) expanding in the appropriated group generators as $\mathcal{D}(\rho) = \mathcal{D}(\rho)_\mu \lambda^\mu = D_{\mu\nu}\rho^\nu \lambda^\mu$, in which $\lambda^{\mu}$ are the generators of SU($N$) for a system of $N$ neutrinos families. In fact, the same procedure can be done in the Hamiltonian term of (\ref{eq:gksl}) in order to get a Lindbladian operator $\mathcal{L} = -2 (H_{\mu\nu} + D_{\mu\nu})$, leading to:

\begin{equation}\label{eq:gksl-liouvillian}
    \ket{\dot{\rho}} = -2 \mathcal{L} \ket{\rho}
\end{equation}
that operates in a ``vectorized" density matrix $\ket{\rho}$ with dimension $N^2$ (where $N$ is the number of levels of the system). In 3 neutrino mixing, $\ket{\rho}$ has dimension 9 and $\mathcal{L}$ is a $9 \times 9$ matrix. 

One of the advantages of this formalism is that, despite a lack of understanding about the microscopic phenomena we are interested to model, we are able to infer the resulting damping effects by properly parameterizing $\mathcal{D}(\rho)$ (or more specifically $D_{\mu\nu}$) in a generic way\footnote{For some forms of $\mathcal{D}(\rho)$ derived from first principles, see~\cite{nieves2019neutrino,nieves2020neutrino}.}

\begin{equation}
\label{eq:D-matrix}
D = \begin{pmatrix}
0 & 0 & 0 & 0 & 0 & 0 & 0 & 0 & 0 \\ 
0 & -\gamma_{1} & \beta_{12} & \beta_{13} & \beta_{14} & \beta_{15} & \beta_{16} & \beta_{17} & \beta_{18} \\ 
0 & \beta_{12} & -\gamma_{2} & \beta_{23} & \beta_{24} & \beta_{25} & \beta_{26} & \beta_{27} & \beta_{28} \\ 
0 & \beta_{13} & \beta_{23} & -\gamma_{3} & \beta_{34} & \beta_{35} & \beta_{36} & \beta_{37} & \beta_{38} \\ 
0 & \beta_{14} & \beta_{24} & \beta_{34} & -\gamma_{4} & \beta_{45} & \beta_{46} & \beta_{47} & \beta_{48} \\ 
0 & \beta_{15} & \beta_{25} & \beta_{35} & \beta_{45} & -\gamma_{5} & \beta_{56} & \beta_{57} & \beta_{58} \\ 
0 & \beta_{16} & \beta_{26} & \beta_{36} & \beta_{46} & \beta_{56} & -\gamma_{6} & \beta_{67} & \beta_{68} \\ 
0 & \beta_{17} & \beta_{27} & \beta_{37} & \beta_{47} & \beta_{57} & \beta_{67} & -\gamma_{7} & \beta_{78} \\ 
0 & \beta_{18} & \beta_{28} & \beta_{38} & \beta_{48} & \beta_{58} & \beta_{68} & \beta_{78} & -\gamma_{8} \\ 
\end{pmatrix} ,
\end{equation}
in 3 neutrino mixing. Although it is not explicit, the entries in matrix (\ref{eq:D-matrix}) can be directly related to the coefficients of expansion of $V_p$ in the generators of SU(3), or $\gamma,\beta = f(v_{p})$, with $v_p$ coming from $V_p = {v_{p}}_\mu \lambda^\mu$. Note that the null entries in the first column of (\ref{eq:D-matrix}) are given by the hermiticity of $V_p$, which also enables rewriting the dissipation term as $\mathcal{D}(\rho) = \frac{1}{2} \sum_p^{N^2-1} [[V_p, \rho], V_{p}]$, showing that terms proportional to identity in the SU(3) expansion vanish, making the first line of (\ref{eq:D-matrix}) also null. It is important to note that the parameters used to define $D_{\mu\nu}$ are not all independent. They are related to each other in order to ensure complete positivity, which is a necessary condition for a quantum state to be physically realizable \cite{benatti1997completely, benatti2000open, carrasco2019probing} (see \cite{carrasco2019probing} for a set of relations in a 3-level system).

However, it is not viable to investigate this general format of (\ref{eq:D-matrix}) given the number of parameters. Therefore, in this work, we restrict ourselves to cases in which $D$ is diagonal as in \cite{de2020solar}, in order to capture the effects of interest arising from QD. We  tested a non-diagonal version of $D$ using complete positivity relations and our results are not significantly affected.

In the context of supernova neutrinos, the neutrino propagates a large distance inside the supernova 
($\sim 10^8$ km), 
then we also investigate the impact of QD combined with SN matter effects. A possible procedure to cross-check it is by rotating eq.~(\ref{eq:gksl}) to flavor basis, where the Hamiltonian can be summed to an MSW potential, i.e. $H_f = H_f^\text{vac} + V_W$. However, as it will be more clear in Section~\ref{sec:factor}, the probability we are interested in is between mass eigenstates on both matter and vacuum, which can be accomplished by diagonalizing the Hamiltonian in flavor basis using a proper transformation. 

\subsection{Selected Models}\label{sec:selected}

Since we analyse diagonal versions of (\ref{eq:D-matrix}), $\beta_{\mu\omega} = 0$ for all $\mu$ and $\omega$. In works such as \cite{oliveira2010quantum, de2020solar} it is shown that quantum decoherence can give rise to two disentangled effects when the evolution occurs in vacuum: the pure decoherence, where a coherent state becomes incoherent along propagation; and the relaxation effect, responsible to lead the ensemble to a maximal mixing. As decoherence effects on SN neutrinos are suppressed due to matter effects on the mixing angle and long propagation lengths\footnote{If neutrinos are only affected by MSW effect, it is possible for $\nu_\mu$ and $\nu_\tau$ oscillate to each other. It generally does not affect the analysis of flavor conversion, once they are indistinguishable in the detection, and therefore generally denoted as $\nu_x$. However, as we will see in Section~\ref{sec:metho-sim}, their creation in coherent states changes one of the tested QD models here.}, we do not expect pure decoherence effects to play any role in the propagation, being only (possibly) affected by relaxation.

\begin{figure}
    \centering
    \includegraphics[width=0.5\textwidth]{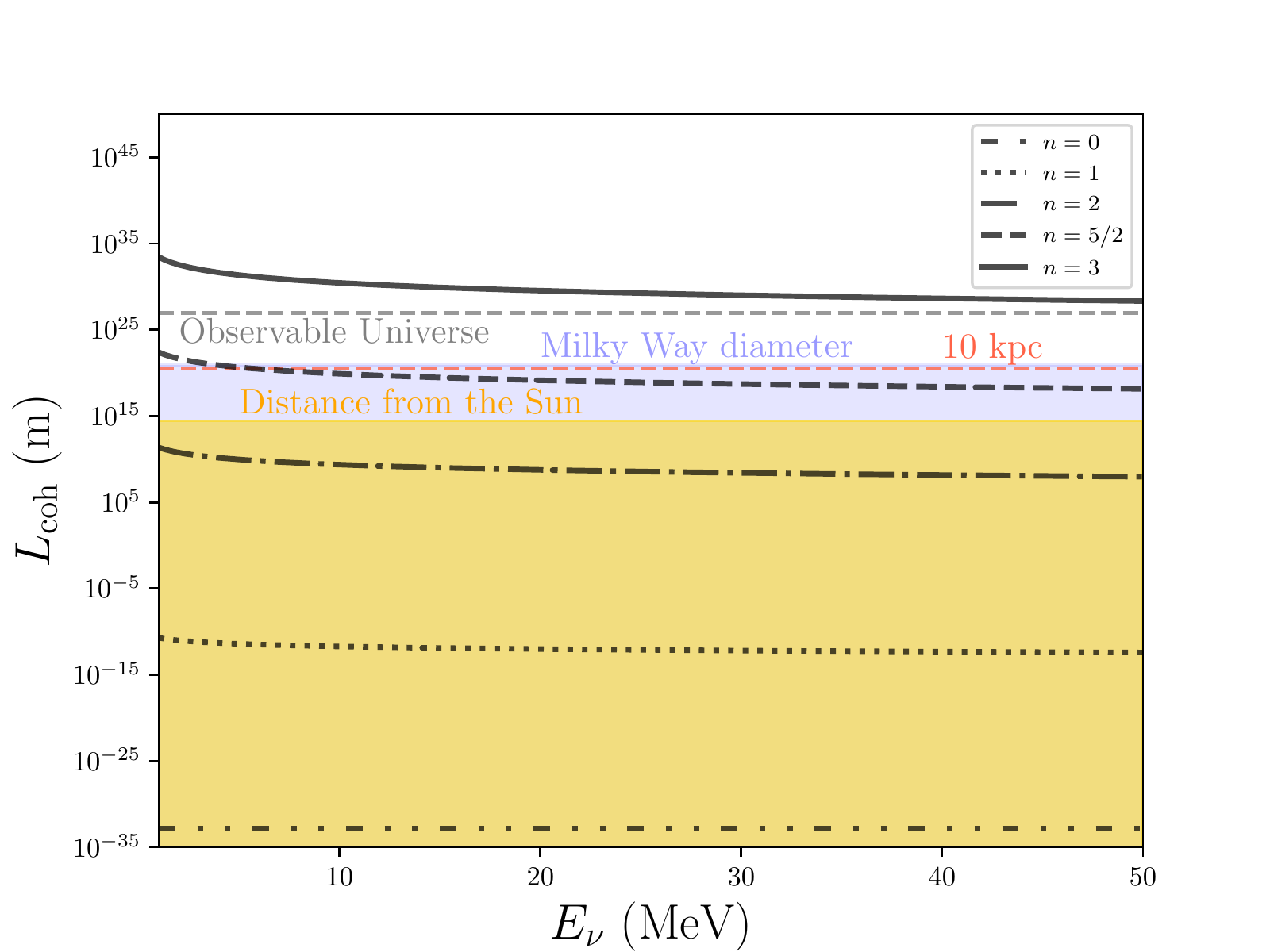}
    \caption{Coherence length ($L_\text{coh} =1/\gamma$) for values of $n$ in a power law of decoherence coefficients $\gamma = \gamma_0 (E/E_0)^n$ for a ``natural" scale of quantum gravity, with $\xi_\text{Planck} = 1$. The yellow region corresponds to the solar system edge, while the blue region is the Milky Way diameter, and the dashed grey line is to respect the observable universe.}
    \label{fig:Lcoh}
\end{figure}

Up to this date, and to the authors's best knowledge, there is no consistent theory in which you can get the parameters of $D$ from quantum gravity, or even if the parameters are constant. Different works \cite{lisi2000probing,amelino2013quantum, coloma2018decoherence,stuttard2020neutrino} suggested the possibility of a dependency on energy as $\gamma_i = \gamma_{0_i} (E/E_0)^n$ motivated by quantum space-time phenomenology, where $E_0$ is an arbitrary energy scale. In this work, we chose $E_0=10$ MeV to match the energy scale of supernova neutrinos. As for the energy dependence, we explore the scenarios with $n = 0$ and $n = 2$, given that most of the works check this power law exponents for $\gamma_i$, which enables us to compare SN limits to other sources (and works), and $n = 5/2$, well-motivated by the natural Planck scale for the SN energy range of $0-100$ MeV. By natural scale, we refer to $\gamma_{0_i} = \xi_\text{Planck}/M_\text{Planck}^{n-1}$ with $\xi_\text{Planck} \sim 1$ \cite{anchordoqui2005probing, stuttard2020neutrino}, making $\gamma_{0i} = \xi_\text{Planck} M_\text{Planck}^{1}$, $\xi_\text{Planck} M_\text{Planck}^{-1}$, and $\xi_\text{Planck} M_\text{Planck}^{-3/2}$ for our choices of $n=0,2$ and 5/2. 

With dimensional analysis (which can be further justified when solving the evolution equation), we expect that the effects of decoherence would show up for distances larger than a {\it coherence length}, defined by $L_\text{coh}=1/\gamma$. 
In Fig.~\ref{fig:Lcoh} we show the expected coherence length for these values of $n$. We see that if this ``natural" scale holds, $n=0$ and 2 would be possibly ruled out by terrestrial and solar experiments, whereas for $n=3$, $L_\text{coh}$ is out of the observable universe for the expected SN-$\nu$ energy scale. For the mentioned values of $n$, we analyze the following models:

\textbf{Mass State Coupling (MSC)}: The neutrino mass basis is coupled to the environment and the relaxation effect leads to maximal mixing. In 3-$\nu$ mixing, it means a 1/3 (equal) probability of detecting any state. In this model, we test two possible scenarios related to energy conservation in the neutrino subsystem:

\begin{itemize}
    \item[$i$)] MSC$^\slashed{\epsilon}$ ($[H,V_p] = 0$): Here, the neutrino energy is conserved for any non-standard mixing process in vacuum\footnote{In our notation, the superscript symbol $\slashed{\epsilon}$ accounts to no exchange of energy to the environment, while $\epsilon$ has the opposite meaning.}. It means that $V_p = \mathbf{v}_3 \lambda_3 + \mathbf{v}_8 \lambda_8$, where $\lambda_\mu$ are Gell-Mann matrices and  $\mathbf{v}_\mu = \sum_{p=1}^8 {v_{p}}_\mu$, with $\mu$ ranging from 0 to 8 in the SU(3) expansion of $V_p$. To simplify the analysis we choose a diagonal version of the dissipation term in (\ref{eq:D-matrix}) with a single parameter $\Gamma$. Additionally, using complete positivity relations \cite{carrasco2019probing}, we can find the special case of $D = -\text{diag}(0,\Gamma,\Gamma,0,\Gamma/4,\Gamma/4,\Gamma/4,\Gamma/4,0)$, with $\Gamma=\Gamma_0 (E/10~{\rm MeV})^n$. The transition probabilities amongst mass states in vacuum are null in this case. However, if we look at the propagation inside the supernova layers, in a diagonalized basis of the mass state in matter $P_{ij}^{m (\text{SN})}$, this probability could be non-null for $i \neq j$, i.e. transitions between $\nu_i^m$ and $\nu_j^m$ are allowed and would change proportionally to $e^{-\Gamma}$. Therefore, the coherence length to be investigated is the SN radius, and the matter effects in addition to quantum decoherence would induce a maximal mixing inside the SN. In Fig.~\ref{fig:Pii-conserved-E} we show the transition probabilities of mass state in matter basis calculated using the slab approach with a simulated SN density profile from Garching group \cite{Garching, serpico2012probing}, corresponding to a progenitor of 40~$M_\odot$. More details about our solution are in Appendix~\ref{appendix:a}. When the neutrino is released to vacuum, it is no longer affected by quantum decoherence until detection. Since the length traveled inside the Earth by the neutrino is much smaller then $L_\text{coh}^\text{SN}$, we do not take the quantum decoherence in Earth matter into account in this specific case, albeit standard non-adiabatic MSW effect could play a role. Note that this regime essentially depends on $\nu$ matter effects in the SN.

\begin{figure}[h!]
    \centering
    \includegraphics[width=0.4\textwidth]{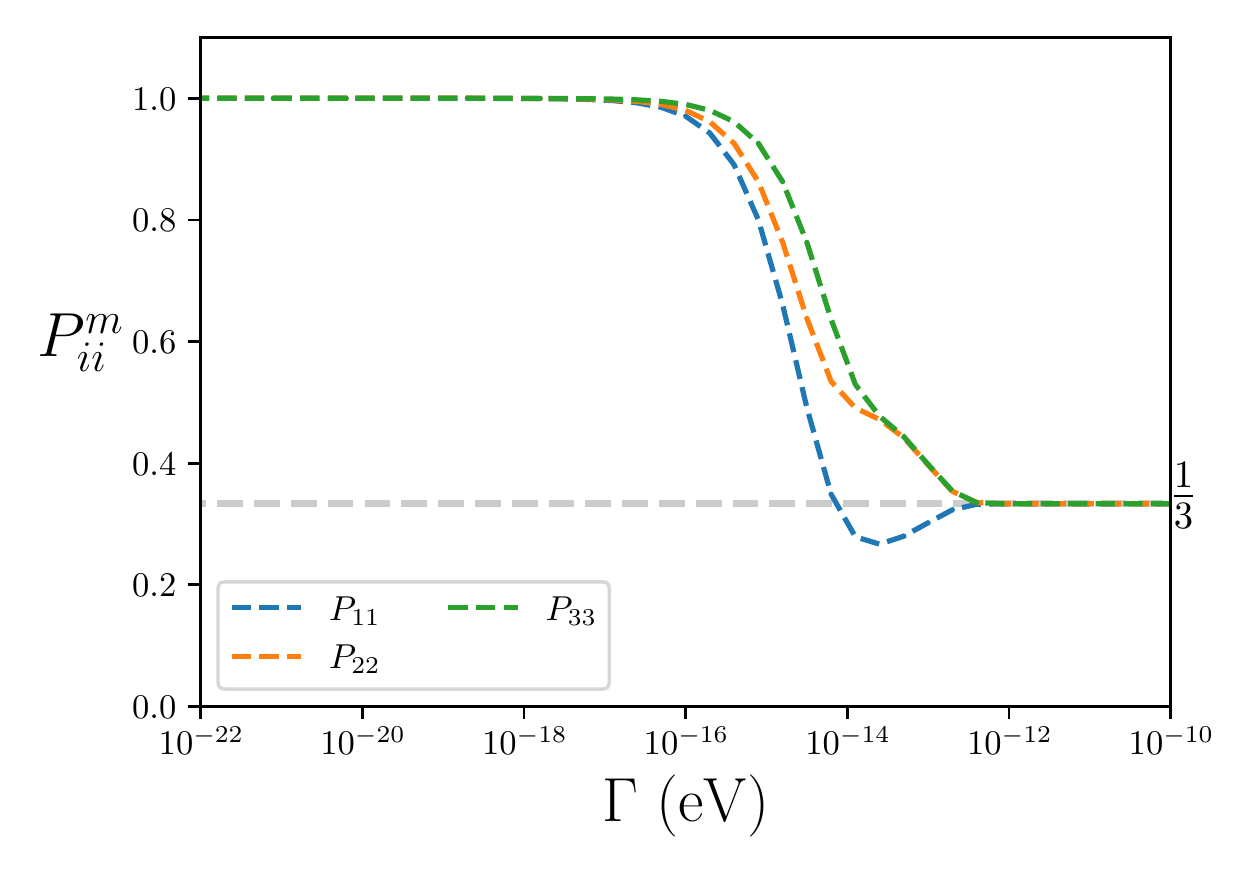}
    \caption{Survival probabilities of mass state in matter basis inside the SN  for the MSC$^{\slashed{\epsilon}}$ model (no exchange of energy from neutrinos and environment in vacuum) and $n=0$ (and then $\Gamma=\Gamma_0$). The SN matter density profile used is from a Garching simulation of a 40~$M_\odot$ (LS180-s40.0) progenitor \cite{Garching, serpico2012probing}, shown in Fig.~\ref{fig:density-sn} in Appendix~\ref{appendix:a}. }
    \label{fig:Pii-conserved-E}
\end{figure}

\item[$ii$)] MSC$^\epsilon$ ($[H,V_p] \neq 0$): In this model, we relax the above assumption, allowing some exchange of $\nu$ energy with the ``non-standard" environment. We choose the most general diagonal version  to the dissipation term from (\ref{eq:D-matrix}): $D=-\text{diag}(0,\gamma_1,\gamma_2,\gamma_3,\gamma_4,\gamma_5,\gamma_6,\gamma_7,\gamma_8)$.
In \cite{stuttard2020neutrino}, this choice of $D$ is intrinsically related to \textit{mass state selected} scenario to be impacted by quantum gravitational effects. To quantify the effects of this model, we solve analytically (\ref{eq:gksl}) to get the  probabilities of interest in mass basis in vacuum\footnote{The expected  (adiabatic MSW) solution for the probabilities is a Kronecker delta, i.e. $P_{ij} = \delta_{ij}$.}:

\begin{equation}\label{eq:Pij-msc}
\begin{split}
    &P_{11} = \frac{1}{3} + \frac{1}{2} e^{-\gamma_3 x} + \frac{1}{6} e^{-\gamma_8 x} \\
    &P_{12} = \frac{1}{3} - \frac{1}{2} e^{-\gamma_3 x} + \frac{1}{6} e^{-\gamma_8 x} \\
    &P_{13} = \frac{1}{3} - \frac{1}{3} e^{-\gamma_8 x} \\
    &P_{33} = \frac{1}{3} + \frac{2}{3} e^{-\gamma_8 x}
\end{split}
\quad\quad
\begin{split}
    &P_{22} = P_{11} \\
    &P_{23} = P_{13} \,\, ,
\end{split}
\end{equation}
with $x$ as the propagated distance. For other possible probabilities on this basis, we use $P_{ij} = P_{ji}$. It should be noted that on this basis the probabilities depend only on $\gamma_3$ and $\gamma_8$. The reason is that when solving the set of differential equations in (\ref{eq:gksl-liouvillian}), the equations corresponding to $\gamma_3$ and $\gamma_8$, i.e. $\mathcal{L}_{3\nu}$ and $\mathcal{L}_{8\nu}$ are the only decoupled ones, independent of Hamiltonian terms.

If we look at $\gamma_i$ parameters in terms of $\mathbf{v}_\mu$ coefficients of the SU(3) expanded $V_p$ we find

\begin{equation}\label{eq:open-g3-g8}
\begin{split}
    \gamma_3 &= \mathbf{v}_1^2 + \mathbf{v}_2^2 + \frac{\mathbf{v}_4^2}{4} + \frac{\mathbf{v}_5^2}{4} + \frac{\mathbf{v}_6^2}{4} + \frac{\mathbf{v}_7^2}{4}\\
    \gamma_8 &= \frac{3\mathbf{v}_4^2}{4} + \frac{3\mathbf{v}_5^2}{4} + \frac{3\mathbf{v}_6^2}{4} + \frac{3\mathbf{v}_7^2}{4} \, .
\end{split}
\end{equation}
Equation (\ref{eq:open-g3-g8}) shows that $\gamma_3$ and $\gamma_8$ are not independent. In order to compare our results to solar limits \cite{de2020solar}, we can use the same notation to define:

\begin{equation}\label{eq:g3,g8-definition}
\begin{split}
    \Gamma_3 &= \mathbf{v}_1^2 + \mathbf{v}_2^2 \\
    \Gamma_8 &= \frac{3\mathbf{v}_4^2}{4} + \frac{3\mathbf{v}_5^2}{4} + \frac{3\mathbf{v}_6^2}{4} + \frac{3\mathbf{v}_7^2}{4}
\end{split}
\end{equation}
leading to $\gamma_3 = \Gamma_3 + \Gamma_8/3$ and $\gamma_8 = \Gamma_8$, resulting in pure (independent) relaxation $\Gamma$ parameters, that will be the ones effectively inducing the maximal admixture in this scenario. The energy dependence is explicitly written as $\Gamma_i=\Gamma_{0i}(E/10~{\rm MeV})^n$ with $i=\{3,8\}$. Note that the effective distance of this particular case is the total neutrino propagation, i.e. vacuum propagation is also affected and it can be split into the regime in the SN and outside its surface until Earth, or $L = L^\text{SN} + L^\text{Vac}$. Similarly as in $i$), we solve the probabilities associated with possible transitions in supernova layers only numerically. However, as we discuss in Section~\ref{sec:factor}, given that $L^\text{Vac} \gg L^\text{SN}$, the approximation of $L \sim L^\text{Vac}$ is assumed in our calculations. 
\end{itemize}

\textbf{Neutrino Loss}: As mentioned in \cite{stuttard2020neutrino}, it is possible to have a scenario with neutrino loss, where neutrinos are captured by effects of quantum gravity during propagation, and re-emitted to a different direction, never reaching the detector at Earth. In this picture, the authors made a choice of $D_{00} \neq 0$. Looking at the most general form of $\mathcal{D}(\rho)$, it is possible to say that this choice is completely out of open quantum systems formalism, i.e. naturally $\mathcal{D}(\rho)_{0\mu} = 0$ when the master equation (\ref{eq:gksl}) is assumed to describe the evolution of the reduced quantum system, with trace-preserving all times. Even though, to explore such an interesting physical situation, we test this non-unitary case that matches the choice $\gamma_i = \gamma$ with $i$ from 1 to 8, then $D = -\text{diag}(\gamma,\gamma,\gamma,\gamma,\gamma,\gamma,\gamma,\gamma,\gamma)$, with $\gamma=\gamma_0(E/10~{\rm MeV})^n$. The solution of (\ref{eq:gksl}) gives:

\begin{equation}\label{eq:Pij-loss}
\begin{split}
    &P_{ii} = e^{-\gamma x} \\
    &P_{ij} = 0
\end{split}
\end{equation}
for any $i,j$ from 1 to 3 with $i \neq j$. Note that in this result, in contradiction to conventional unitary models, one state does not go to another, i.e. $\sum_i P_{ij} \neq 1$, once neutrinos are lost along the way.

In the solutions of the equation of motion shown above, we absorbed a factor of 2 in the quantum decoherence parameters, i.e. $-2\gamma_i \rightarrow -\gamma_i$, with no loss of generality, since what matters in our results is the intensity of a deviation from a standard scenario.

\section{Methodology and simulation}\label{sec:metho-sim}

To test the QD models discussed in the context of a future SN detection, we use the neutrino flux coming from supernovae simulations from the Garching group \cite{Garching}. For MSC$^{\slashed{\epsilon}}$ described in item $i$) of MSC in Section~\ref{sec:selected}, we exploit a 40~$M_\odot$ progenitor simulation (LS180-s40.0) \cite{serpico2012probing}, since it has detailed matter density profiles, essential to explore such scenario. For all other cases investigated (MSC$^\epsilon$ and $\nu$-loss), we use simulations with 27~$M_\odot$ (LS220s27.0c) and 11.2~$M_\odot$ (LS220s11.2c) progenitor stars, detailed in \cite{mirizzi2016supernova}.

\begin{figure}[h]
    \centering
    \includegraphics[width=0.48\textwidth, height=0.346\textwidth]{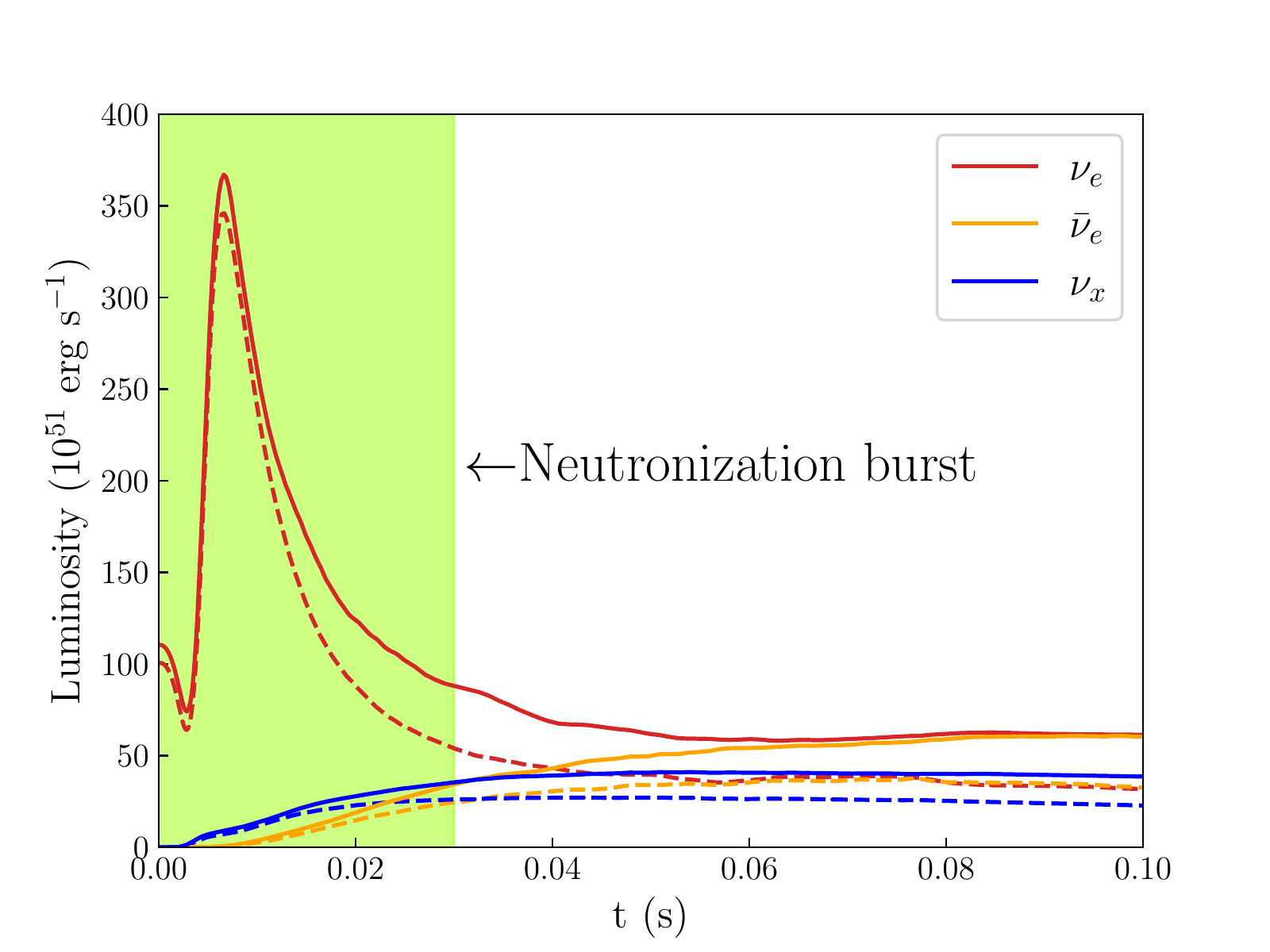}
    \caption{Simulated $\nu$ luminosity for neutronization burst phase of the emission models of 27 $M_\odot$ (solid) and 11.2 $M_\odot$ (dashed)  progenitor stars from Garching group \cite{Garching,mirizzi2016supernova}.}
    \label{fig:lumi}
\end{figure}

To avoid the large uncertainties of collective effects, we only use the flux from the neutronization burst phase (first 30 ms) in our analysis, in which effects induced by $\nu-\nu$ interaction are expected to not play a significant role. In Fig.~\ref{fig:lumi} we show the luminosity of all flavors along the time window of this phase. 

Next, we explain in more detail how to include non-standard physics of eqs.~(\ref{eq:Pij-msc}) and (\ref{eq:Pij-loss}) in SN neutrino evolution and our methods to use a future SN detection to impose limits on QD parameters.

\subsection{Factorization of the dynamics}
\label{sec:factor}

Our analysis only takes into account the MSW effect in the neutronization burst through the standard matter effect on $\nu$ mixing. To combine QD effects and MSW through the $\nu$ generation, propagation, crossing through Earth, and detection, it is possible to factorize the flavor probabilities as 

\begin{equation}\label{eq:pee-factorized}
   P_{\alpha\beta} = \sum_{i,j,k = 1}^3 P_{\alpha i}^{m (\text{SN})} P_{ij}^{m (\text{SN})} P_{jk} P_{k\beta}^{m (\text{Earth})} \quad\quad\quad \bar{P}_{\alpha\beta} = \sum_{i,j,k = 1}^3 \bar{P}_{\alpha i}^{m (\text{SN})} \bar{P}_{ij}^{m (\text{SN})} \bar{P}_{jk} \bar{P}_{k\beta}^{m (\text{Earth})} \,\,,
\end{equation}
where $P_{\alpha\beta}$ ($\bar{P}_{\alpha\beta}$) are the transition probabilities from flavor $\alpha$ to $\beta$. The meaning of each term in  (\ref{eq:pee-factorized}) can be summarized as: $P_{\alpha i}^{m (\text{SN})}$ is the probability of creating a $\nu_\alpha$ as a $i$ state in matter $\nu_{i}^m$; $P_{ij}^{m (\text{SN})}$ is the probability of converting $\nu_{i}^m \rightarrow \nu_{j}^m$ inside supernova layers; $P_{jk}$ the probability of converting $\nu_j \rightarrow \nu_k$ during propagation in vacuum until Earth; and by the end, $P_{k\beta}^{m (\text{Earth})}$ is the probability of detecting a $\nu_\beta$ given a $\nu_k$ state considering (or not) Earth matter effects. The index $m$ regards that the creation or propagation is in matter. It is worth remembering that $\nu_e$ and $\bar{\nu}_e$ are created as a single mass eigenstate in matter. In this scenario, the sum over $i$ vanishes, since we have $P^{m (\text{SN})}_{ei}=\delta_{i3}$ and $\bar P^{m (\text{SN})}_{ei}=\delta_{i1}$ for NH, and $P^{m (\text{SN})}_{ei}=\delta_{i2}$ and $\bar P^{m (\text{SN})}_{ei}=\delta_{i3}$ for IH. 
As for $\nu_x$, although it is created in a coherent superposition of the other two mass eigenstates, the interference phase would be averaged out, and therefore eq.~(\ref{eq:pee-factorized}) is valid. In the context of a SN flux conservation, the simplest flavor conversion scheme could be described by just $P_{ee}$ and $\bar{P}_{ee}$, and in standard neutrino mixing, the factorized probabilities in (\ref{eq:pee-factorized}) become $P_{ij}^{m (\text{SN})} = \delta_{ij}$, $P_{jk} = \delta_{jk}$ and $\bar{P}_{ij}^{m (\text{SN})} = \delta_{ij}$, $\bar{P}_{jk} = \delta_{jk}$ for adiabatic evolution. Such a scenario can be changed by quantum decoherence, allowing for the conversion among mass eigenstates in vacuum and matter.

One can also note in (\ref{eq:Pij-msc}), (\ref{eq:open-g3-g8}), (\ref{eq:g3,g8-definition}) and (\ref{eq:pee-factorized}) that for the MSC$^{\epsilon}$ model, $P_{ee}$ is a function of $\Gamma_3$ and $\Gamma_8$ in IH but only of $\Gamma_8$ for NH. The $\bar{P}_{ee}$ has the opposite dependency and we can write:
\begin{eqnarray*}
    P_{ee}^\text{IH} = P_{ee}^\text{IH}(\Gamma_3, \Gamma_8)~~~&;&~~~P_{ee}^\text{NH} = P_{ee}^\text{NH}(\Gamma_8).
\\
\bar{P}_{ee}^\text{IH} = \bar{P}_{ee}^\text{IH}(\Gamma_3)~~~&;&~~~\bar{P}_{ee}^\text{NH} = \bar{P}_{ee}^\text{NH}(\Gamma_3, \Gamma_8) .
\end{eqnarray*} 
These remarks on the survival probabilities of $\nu_e$ and $\bar{\nu}_e$ are essential in our results, once the flavor conversion of MSC can be described using uniquely $P_{ee}$ and $\bar{P}_{ee}$.

Particularly for the MSC$^\slashed{\epsilon}$ case, considering the propagation along supernova layers, $P_{ij}^{m(\text{SN})}$ and $\bar{P}_{ij}^{m(\text{SN})}$ will be affected by QD, nevertheless $P_{jk} = \delta_{jk}$ and $\bar{P}_{jk} = \delta_{jk}$, since with no exchange of energy to the environment, quantum decoherence would not play any role in the vacuum propagation. On the other hand, for MSC$^\epsilon$, both SN matter and vacuum would affect the neutrino mixing. However, as shown in Fig.~\ref{fig:P33sn} in the Appendix~\ref{appendix:a}, it would be needed a $\Gamma_{3,8} \gtrsim 10^{-18}$ eV or even beyond to have significant changes over $P_{ij}^{m(\text{SN})}$. As it will be clear in Section~\ref{sec:future-limits}, this value is much higher than the possible sensitivity of a future SN detection with only vacuum effects (given the large coherence length between the SN and Earth), then we take $P_{ij}^{m(\text{SN})}$ and $\bar{P}_{ij}^{m(\text{SN})}$ as $\delta_{ij}$ for MSC$^\epsilon$ from now on. 

In order to put bounds on QD effects, we statistically analyze it in two scenarios: without Earth matter effects in neutrino (antineutrino) propagation, or $P_{ke}^{m (\text{Earth})} = P_{ke}$ ($\bar{P}_{ke}^{m (\text{Earth})} = \bar{P}_{ke}$) in (\ref{eq:pee-factorized}); and then we check how Earth matter effects would impact our results.

\begin{figure}[h]\label{fig:pee-combined}
    \centering
        \includegraphics[width=0.4\textwidth]{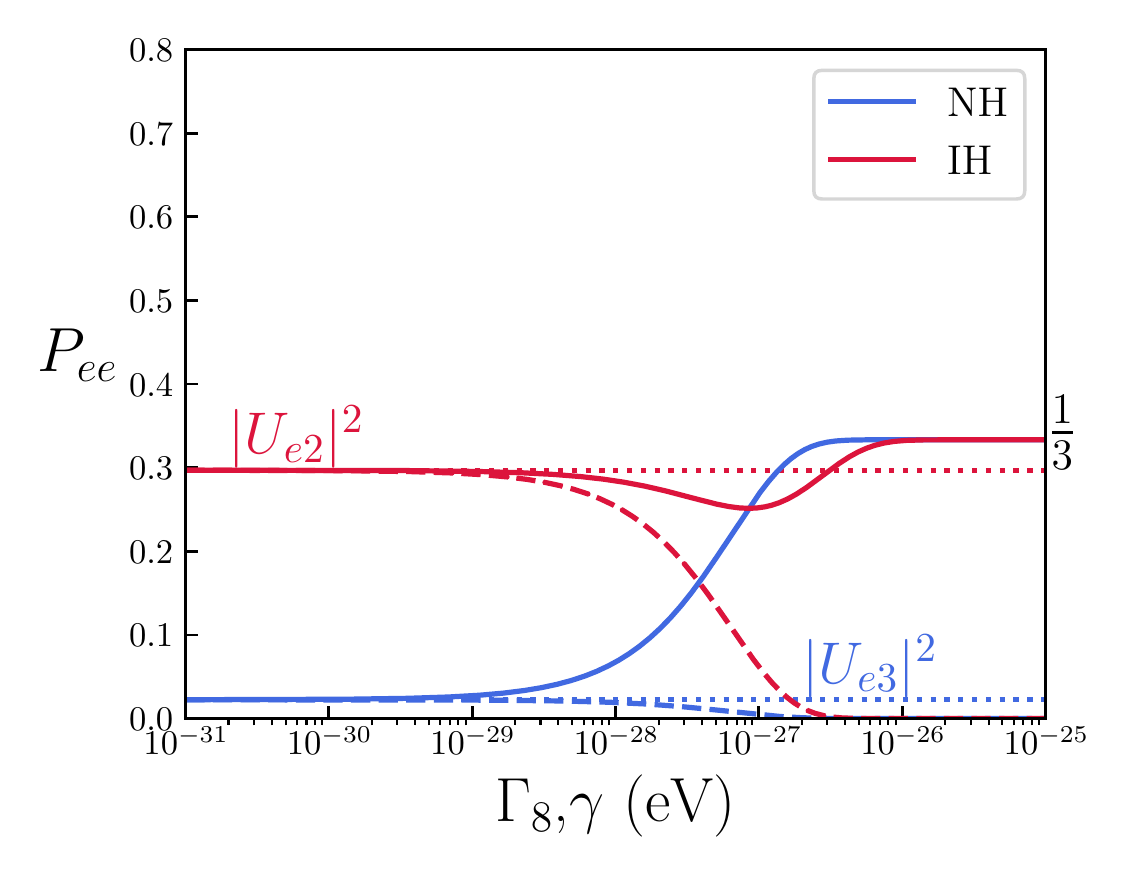}\includegraphics[width=0.4\textwidth]{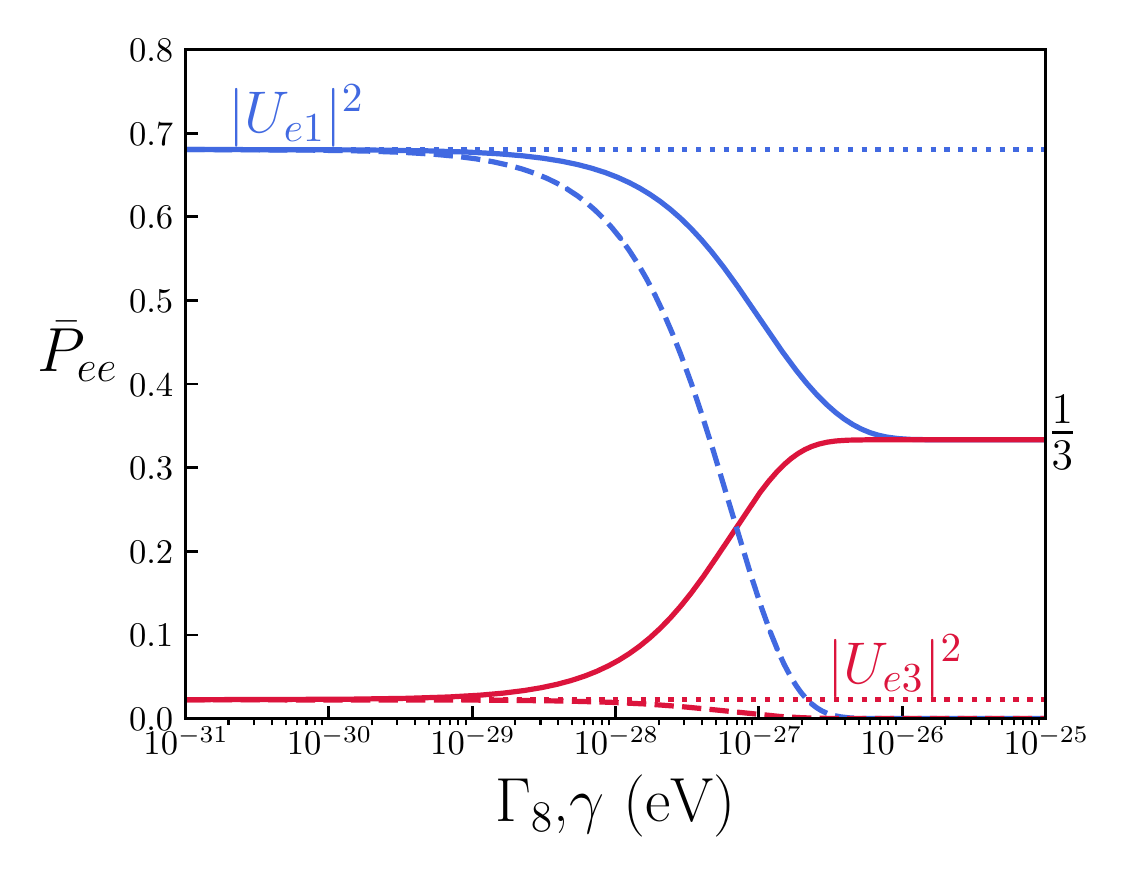}
        \includegraphics[width=0.4\textwidth]{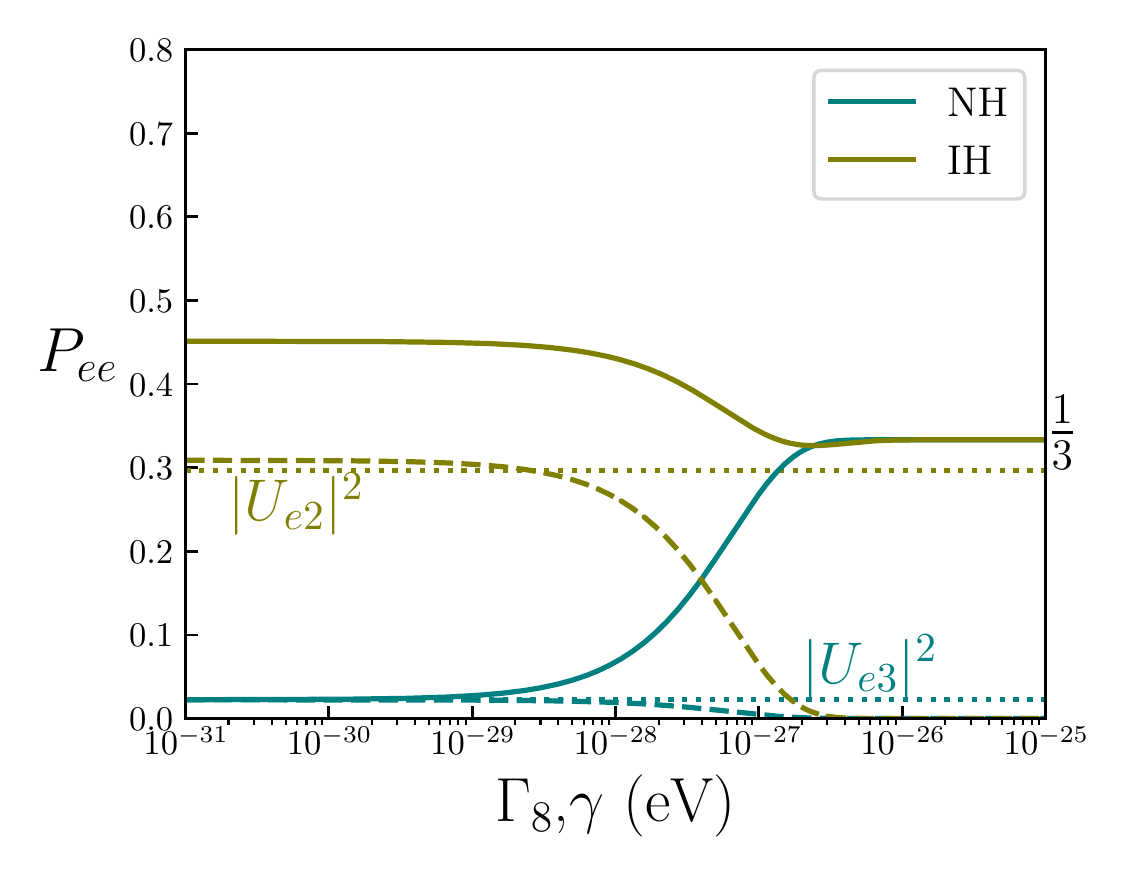}\includegraphics[width=0.4\textwidth]{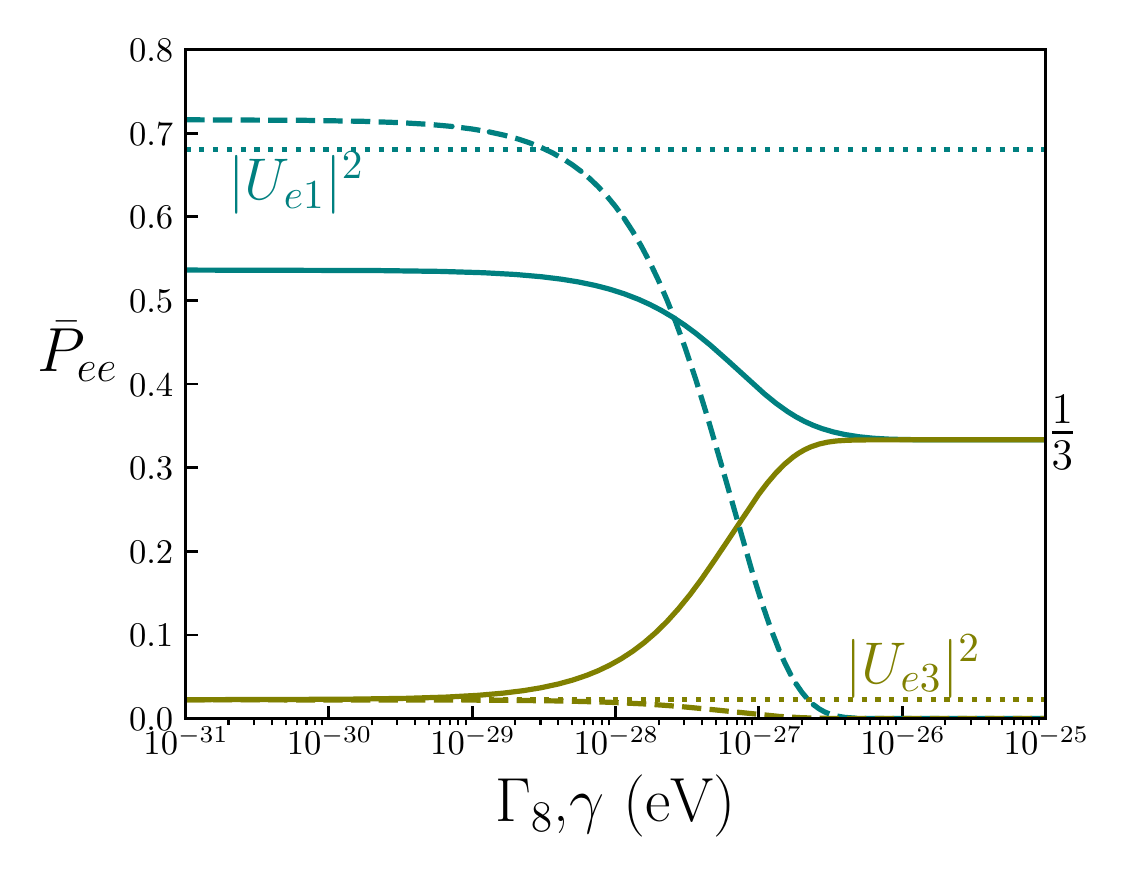}
    \caption{Survivor probability for electron neutrinos (left) and antineutrinos (right) as a function of decoherence parameters for $n = 0$ (energy independent) and a 10~kpc propagation, without (upper plots) and with (down plots) Earth matter effects. Solid lines represent MSC$^\epsilon$ scenario ($\Gamma_8$) with $\Gamma_3 = 10^{-27}$~eV and the dashed, the neutrino loss ($\gamma$). For the upper plots, quantum decoherence is taken into account only in vacuum in between SN surface until detection at Earth, with no regeneration considered. In the down ones, we set the zenith angle of $\theta_z = 180^o$ and $E_\nu = 30$ MeV.}
\end{figure}

Figure \ref{fig:pee-combined} shows both scenarios of $P_{ee}$ and $\bar{P}_{ee}$ as a function of quantum decoherence parameters for neutrinos and antineutrinos, where neutrino hierarchy plays a relevant role in the considered scenarios. It is possible to see that Earth regeneration could enhance or decrease the sensitivity of standard physics on QD parameters for very specific energies and zenith angles $\theta_z$. However, as we will see later, regeneration becomes more relevant for higher energies, generally at the end of the SN-$\nu$ simulated spectrum, limiting its impact on SN flavor conversion.

It is worth mentioning that for the MSC model asymptotically we expect more sensitivity on $P_{ee}$ in NH than IH, since for IH the standard probability is about the maximal admixture (1/3). In contrast, for $\bar{P}_{ee}$, both hierarchy scenarios are almost equally sensitive to a maximal admixture scenario. In the case of $\nu$-loss we see the opposite picture for $P_{ee}$, i.e. IH would be more impacted by an asymptotically null probability, and for $\bar{P}_{ee}$ NH would be highly affected, with low impact on IH.

As we will see later, the most general scheme of SN-$\nu$ fluxes at Earth can not be parameterized with just $P_{ee}$ and $\bar{P}_{ee}$ for the $\nu$-loss scenario, given no conservation of total flux. Therefore it is needed to work out $P_{\alpha\beta}$ also for $\alpha,\beta = \mu,\tau$ (not shown in Figs.~\ref{fig:pee-combined} for simplicity). We clarify it in the next section.

\subsection{Exploring a future SN-$\nu$ detection}
\label{sec:nu-detec}

\begin{figure}
    \centering
    \includegraphics[width=0.4\textwidth]{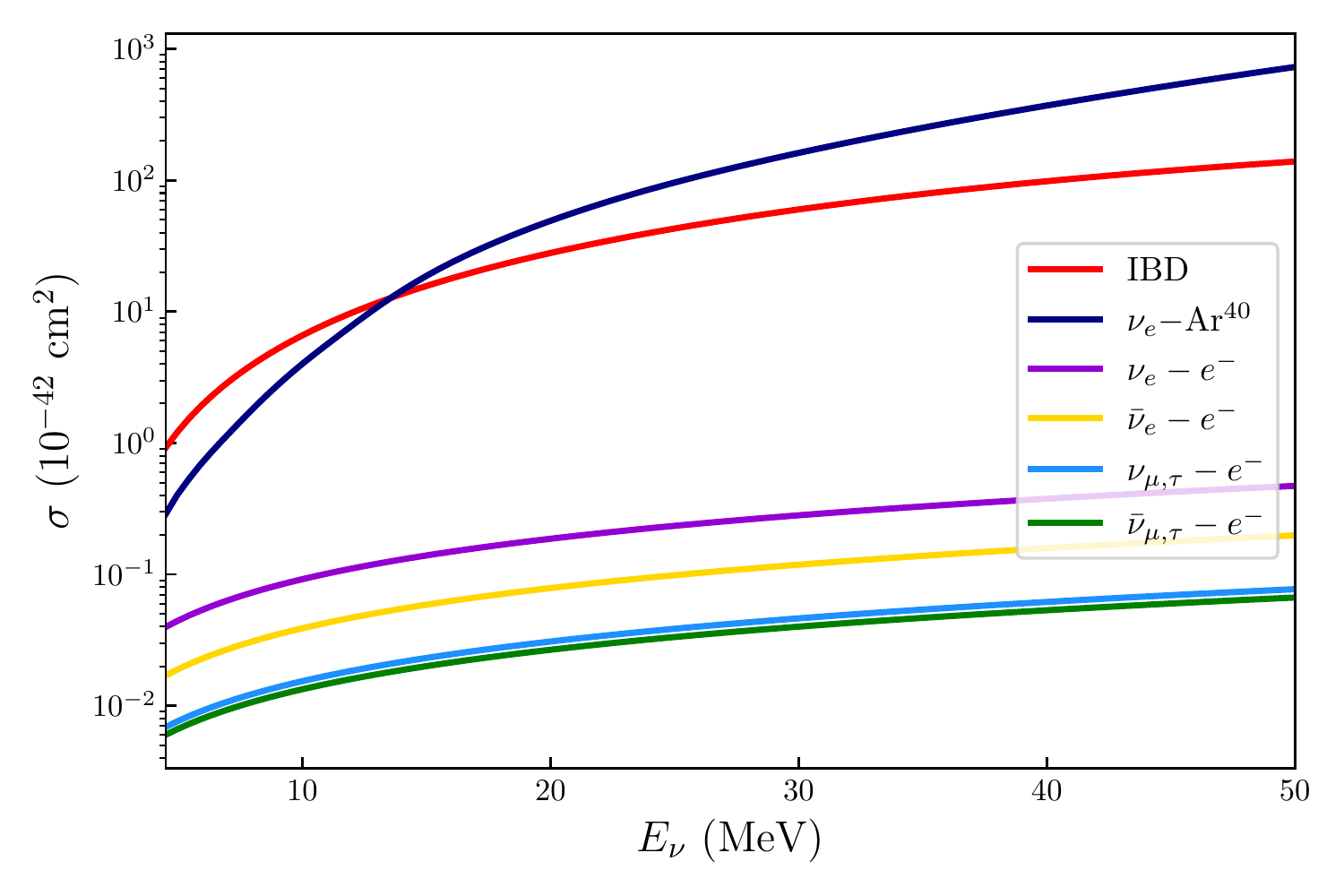}
    \caption{$\nu$ total cross-sections for inverse beta decay (IBD) \cite{vogel1999angular}, $\nu_e -$Ar charge current interaction (from SNOwGLoBES) \cite{scholberg2021snowglobes} and elastic $\nu-e^-$ interaction \cite{de2020measuring}.}
    \label{fig:cross}
\end{figure}

Since the detection of SN1987A through neutrinos, a galactic SN is expected by the community as a powerful natural $\nu$ laboratory. The SN1987A neutrino detection greatly impacted what we know about SN physics, but the low statistics of the available data make predictions on standard $\nu$ admixture extremely challenging. On the other hand, the next generation of neutrino detectors promises a precise measurement of a galactic SN, highly increasing our knowledge of SN-$\nu$ flavor conversion, with different detector technologies and capabilities. Here, we show the sensitivity of DUNE, HK, and JUNO on QD. These detectors have the following properties:

\begin{itemize}
    \item[a)] DUNE will be a 40 kt Liquid-Argon TPC in the USA. We consider only the most promising detection channel $\nu_e + \text{Ar} \rightarrow e^- + K^+$ \cite{abi2021supernova} in our analysis, being sensitive to electron neutrinos and consequently to most neutronization burst flux\footnote{Actually, it depends on the neutrino mass hierarchy, once for MSW-NH the $\nu_e$ flux is highly suppressed.}. We set an energy threshold to $E_\text{th} = 4.5$ MeV and use the most conservative reconstruction efficiency reported in \cite{{abi2021supernova}}. 
    \item[b)] Hyper-Kamiokande will be a water Cherenkov detector in Japan with a fiducial mass of $\sim 374$ kt with main detector channel as the inverse beta decay (IBD), sensible to electron antineutrinos: $\bar{\nu}_e + p \rightarrow e^+ + n$. It is also expected several events from elastic scattering with electrons, with the advantage of sensitivity to all flavors: $\nu + e^- \rightarrow \nu + e^-$. We consider both channels in our analysis. We set a 60\% overall detector efficiency and $E_\text{th} = 3$ MeV.
    \item[c)] JUNO will be a liquid scintillator detector with a fiducial mass of 17 kt situated in China \cite{juno2022juno}. Despite the interesting multi-channel detection technology reported by the collaboration, we take into account only IBD events. We set an overall efficiency of 50\% and $E_\text{th} = 3$ MeV in our analysis.
\end{itemize}

\begin{figure}
    \centering
    \includegraphics[width=0.9\textwidth]{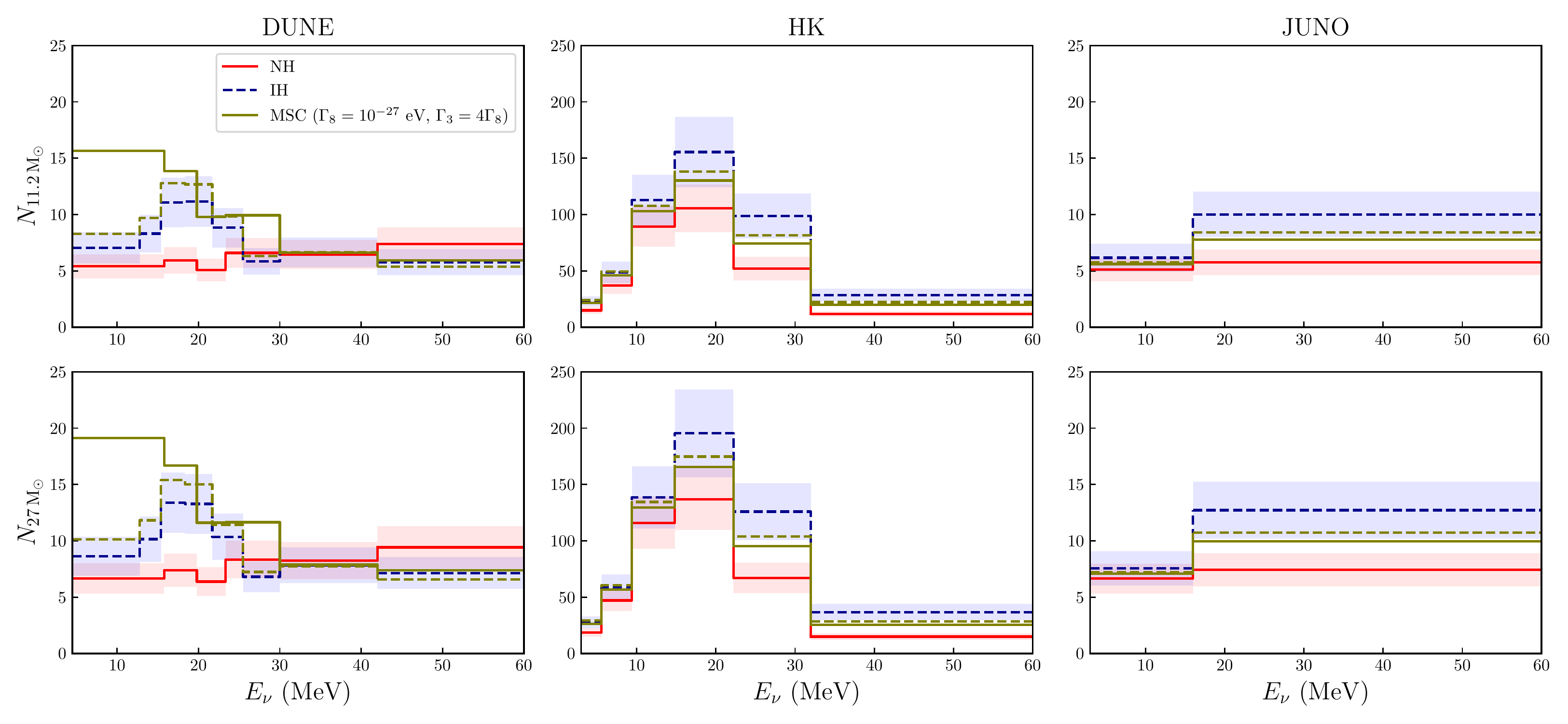}
    \caption{Spectrum of events for DUNE, HK and JUNO for NH (solid lines) and IH (dashed), with $n = 0$ for a 10 kpc SN with 11.2~$M_\odot$ and 27~$M_\odot$ progenitor mass simulations. Each column concerns a detector, while the rows are related to progenitor masses. The size of bins is at least twice the resolution at the specific energy and given a minimum threshold in the number of events per bin established in our analysis. The bands are to respect the 40\% of the uncertainty of the flux over standard NH and IH, with details in the text. For the QD parameters, we used the values $\Gamma_8=10^{-27}$ eV and $\Gamma_3=4\Gamma_8$.}
    \label{fig:events-msc}
\end{figure}

In order to compare the examined scenarios, we will consider only the energy information, calculating the number of events in the $j$-th energy bin as
\begin{equation}\label{eq:events}
    N_j = n_d^c \int_0^\infty dt \int_{0}^{\infty} dE_\nu \frac{d^2\phi_\nu}{dtdE_\nu} \eta(E_\nu) \int_{E_i}^{E_f}  d\bar{E}_\nu R_j(\bar{E}_\nu, E_\nu) \sigma(E_\nu) ,
\end{equation}
where $n_d^c$ is the number of targets for each detector $d$, with $c$ accounting for each specific channel, $\phi_\nu$ is the neutrino flux, $\eta(E_\nu)$ is the efficiency that can eventually depend on $\nu$ energy, $\sigma$ is the neutrino cross-section (with each channel shown in Fig.~\ref{fig:cross}), $R_j$ is the detector resolution. We analyze the $\nu$ energy from the threshold of each detector up to 60~MeV. The $\nu$ mixing is encoded in the flux $\phi_\nu$, that can be written as 

\begin{equation}\label{eq:flux-sn}
\begin{split}
    &\phi_{\nu_e} = \phi_{\nu_e}^0 P_{ee} + \phi_{\nu_x}^0(1 - P_{ee}) \\
    &\phi_{\bar{\nu}_e} = \phi_{\bar{\nu}_e}^0 \bar{P}_{ee} + \phi_{\nu_x}^0(1 - \bar{P}_{ee}) \\
    &\phi_{\nu_x} = \phi_{\nu_e}^0 (1 - P_{ee}) + \phi_{\nu_x}^0 (2 + P_{ee} + \bar{P}_{ee}) + \phi_{\bar{\nu}_e}^0 (1 - \bar{P}_{ee})
\end{split}
\end{equation}
for the standard MSW (widely found in literature, see \cite{Dighe:1999bi,mirizzi2016supernova} for a review), where $\phi_{\nu_\alpha}^0$ refers to initial SN neutrino fluxes and non-standard QD effects are hidden in $P_{ee}$ and $\bar{P}_{ee}$. In Fig.~\ref{fig:events-msc}, the expected number of events for the three detectors are reported in  the energy spectrum of simulated progenitors (11.2 $M_\odot$ and 27 $M_\odot$) for both hierarchies and are compared to MSC$^\epsilon$ model. The results translate what is shown in Fig.~\ref{fig:pee-combined}, weighted by detector capabilities. Expected changes in the spectrum look more prominent when NH is assumed as a standard solution for DUNE, with an increase of $\nu_e$ events for both hierarchies. On the other hand, for HK and JUNO the MSC$^\epsilon$ effect results in a decrease of events in IH and an increase in NH and it is not so clear which hierarchy would be more sensible to the MSC$^\epsilon$ effect, since the number of QD parameters for each one is different for both $P_{ee}$ and $\bar{P}_{ee}$. For instance, for $\bar{P}_{ee}^\text{NH}$, fixing $\Gamma_3$, an increase in $\Gamma_8$ is weighted by the factor 1/3 in the exponential terms, while $\bar{P}_{ee}^\text{IH}$ is more sensible to $\Gamma_8$, since the same change is multiplied by a factor 1, but it is also independent of $\Gamma_3$.

\begin{figure}[h]
    \centering    
    \includegraphics[width=0.8\textwidth]{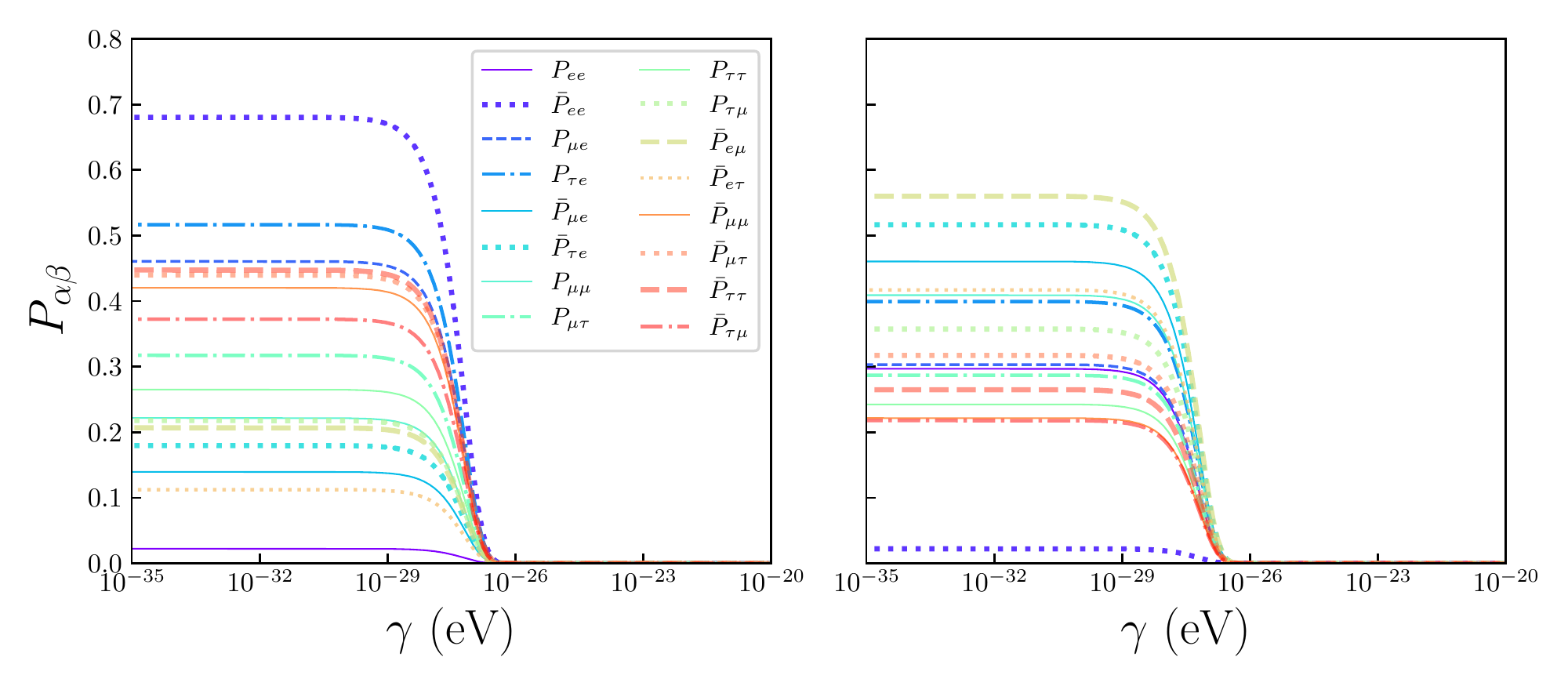}
    \caption{Probabilities with impact of $\nu$-loss with $n = 0$ considering a 10~kpc SN for NH (left) and IH (right). }
    \label{fig:Pab_loss}
\end{figure}

\begin{figure}
    \centering
    \includegraphics[width=0.9\textwidth]{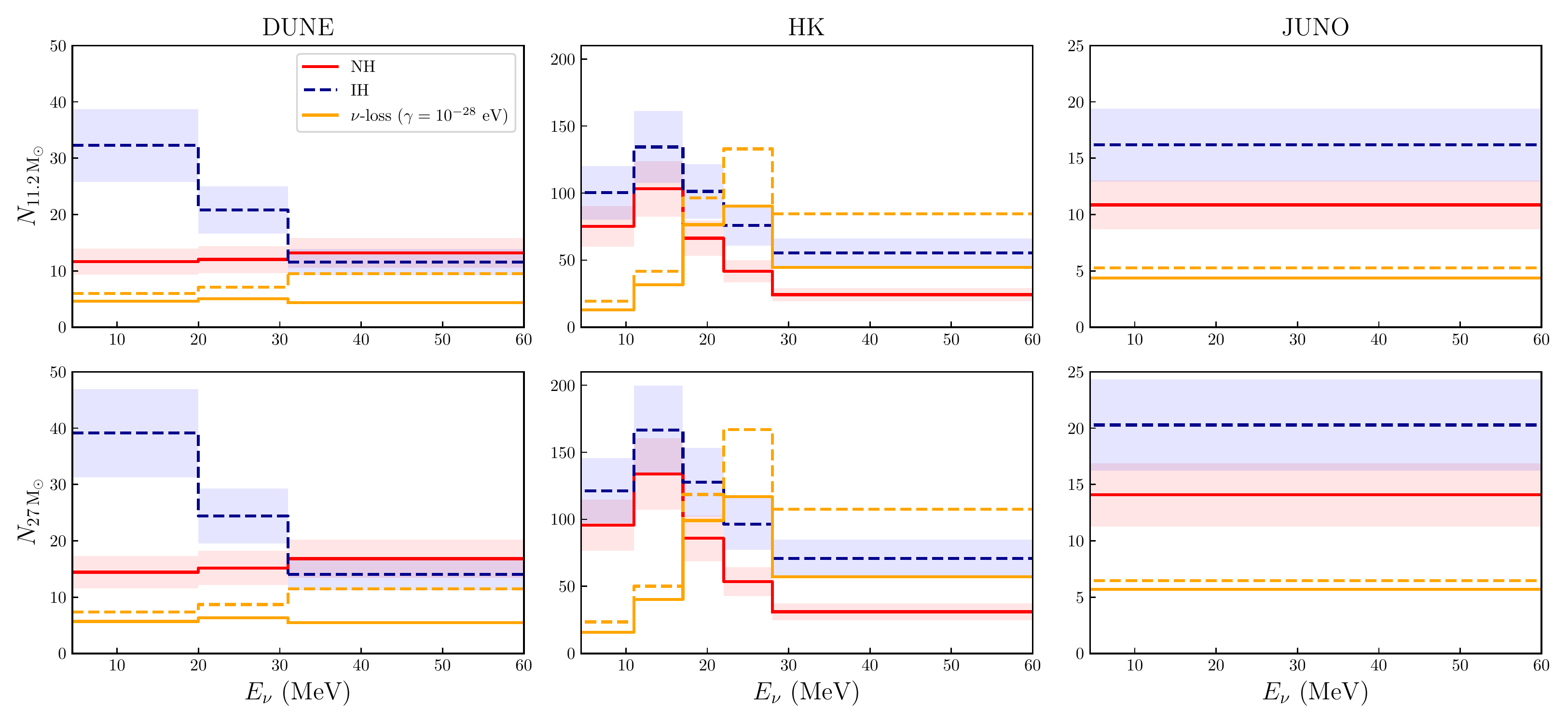}
    \caption{Spectrum of events for DUNE, HK and JUNO for NH (solid lines) and IH (dashed) compared to $\nu$-loss model, with $n = 0$ for a 10 kpc SN with 11.2~$M_\odot$ and 27~$M_\odot$ progenitor mass simulations. For $\nu$-loss we use different bin sizes in order to achieve the requirement of minimum number of events per bin of $\sim 5$. Given the lack of events in this scenario, we decided to use a single bin for JUNO.}
    \label{fig:events-loss}
\end{figure}

Note that eq.~(\ref{eq:flux-sn}) is valid for a conserved total flux, which does not remain in the $\nu$-loss scenario. To get around this issue we propose a more generalized form of (\ref{eq:flux-sn})

\begin{equation}\label{eq:flux-sn-loss}
\begin{split}
    &\phi_{\nu_e} = \phi_{\nu_e}^0 P_{ee} + \phi_{\nu_x}^0(P_{\mu e} + P_{\tau e}) \\
    &\phi_{\bar{\nu}_e} = \phi_{\bar{\nu}_e}^0 \bar{P}_{ee} + \phi_{\nu_x}^0(\bar{P}_{\mu e} + \bar{P}_{\tau e}) \\
    &\phi_{\nu_x}^\prime = \phi_{\nu_e}^0 (P_{e \mu} + P_{e \tau}) + \phi_{\nu_x}^0 (P_{\mu\mu} + P_{\mu\tau} + P_{\tau\tau} + P_{\tau\mu}) \\
    &\phi_{\bar{\nu}_x}^\prime = \phi_{\bar{\nu}_e}^0 (\bar{P}_{e \mu} + \bar{P}_{e \tau}) + \phi_{\bar{\nu}_x}^0 (\bar{P}_{\mu\mu} + \bar{P}_{\mu\tau} + \bar{P}_{\tau\tau} + \bar{P}_{\tau\mu}) \\
    &\phi_{\nu_x} = \phi_{\nu_x}^\prime + \phi_{\bar{\nu}_x}^\prime
\end{split}
\end{equation}
where each probability can be factorized as described in (\ref{eq:pee-factorized}). For the ones where $\alpha = \mu,\tau$, since these flavors are generated in a superposition of mass states in matter, the $\nu_\mu-\nu_\tau$ mixing should be taken into account, where $P_{\alpha i}^{m\text{SN}}$ and $\bar{P}_{\alpha i}^{m\text{SN}}$ would correspond to the proper square module of elements from $U_{\mu\tau}$ mixing matrix\footnote{In the $\mu-\tau$ sector, such probability is associated to $\theta_{23}$ mixing, being a sub-matrix of $U_{23}$ in the conventional PMNS decomposition. We also assume in this formula that any oscillation term is averaged out.}. In Fig.~\ref{fig:Pab_loss} we show each probability $P_{\alpha\beta}$ for a 10~kpc SN for the $\nu$-loss scenario. In Fig.~\ref{fig:events-loss} we show the expected spectrum of events for the $\nu$-loss model.

\subsection{Role of Earth matter effects}

\begin{figure}\label{fig:regeneration-E}
    \centering
        \includegraphics[width=0.8\textwidth]{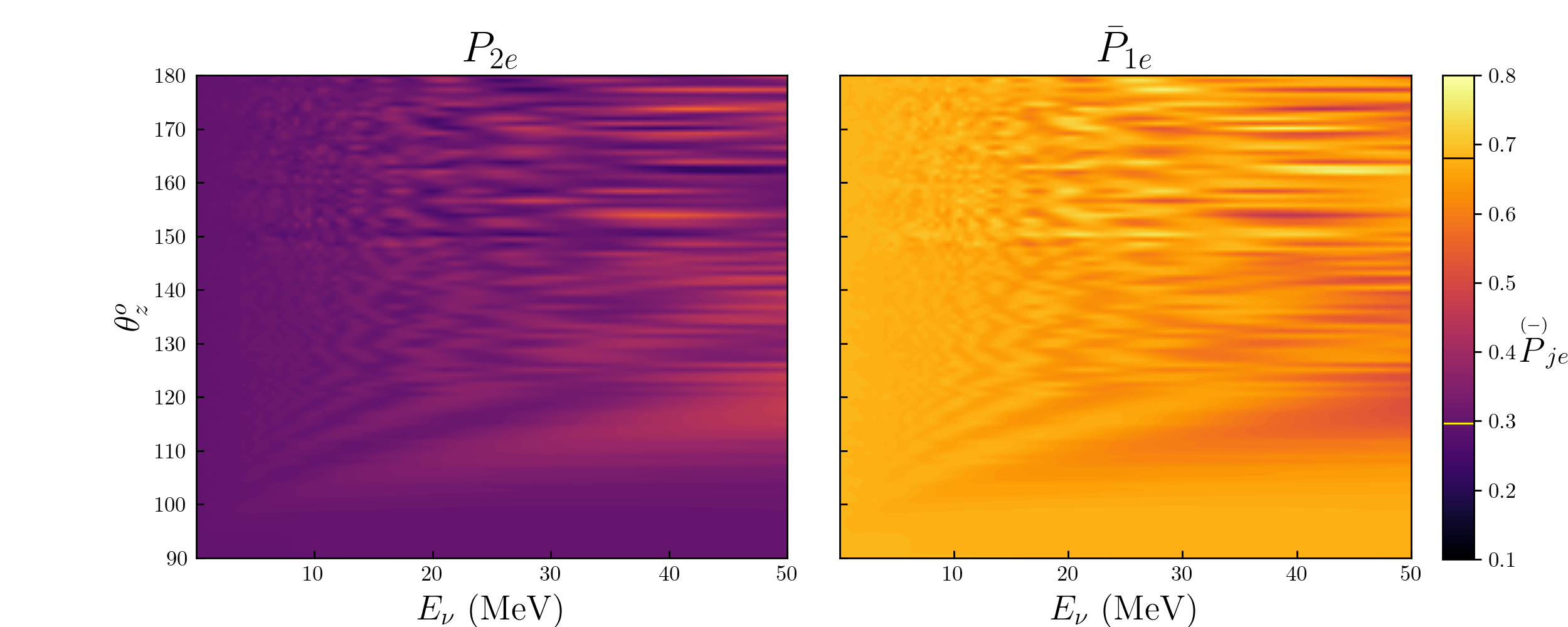}
    \caption{$P_{2e}$ (left) and $\bar{P}_{1e}$ (right) under Earth matter effects as a function of neutrino energy and zenith angle. In standard MSW in supernova mixing, $P_{2e}$ and $\bar{P}_{1e}$ can be used to calculate the survival probabilities of $\nu_e$ (IH) and $\bar{\nu}_e$ (NH) respectively. The lines on the color bar are the adiabatic solutions for $P_{2e}$ (yellow) and $\bar{P}_{1e}$ (black) without regeneration effects.}
\end{figure}

Since a galactic SN detection can be impacted by Earth matter effects, we also calculate $P_{ee}$ and $\bar{P}_{ee}$ to each detector given the position of the SN in the sky. However, as shown in \cite{pompa2022absolute}, it is not expected to play an important role for the neutronization burst. The reason is that regeneration would start to be important beyond $E_\nu \gtrsim 50$ MeV or even higher energies, which is close to the end of the expected spectrum. In Fig.~\ref{fig:regeneration-E} we show the impact of Earth matter effects in $P_{ee}$ for a SN flux of $\nu_e$ in IH and $\bar{P}_{ee}$ for $\bar{\nu}_e$ in NH in a range of zenith angles for only non-adiabatic MSW effect (no quantum decoherence effects) using the PREM density profile available in \cite{anderson1989theory}, where $90^o$ is a horizon of an observer at Earth (with no matter effects) and $180^o$ represents a propagation all along Earth diameter. Note that for $P_{ee}$ in NH and $\bar{P}_{ee}$ in IH, regeneration does not play an important role. 

\begin{figure}\label{fig:regeneration-g}
        \includegraphics[width=0.8\textwidth]{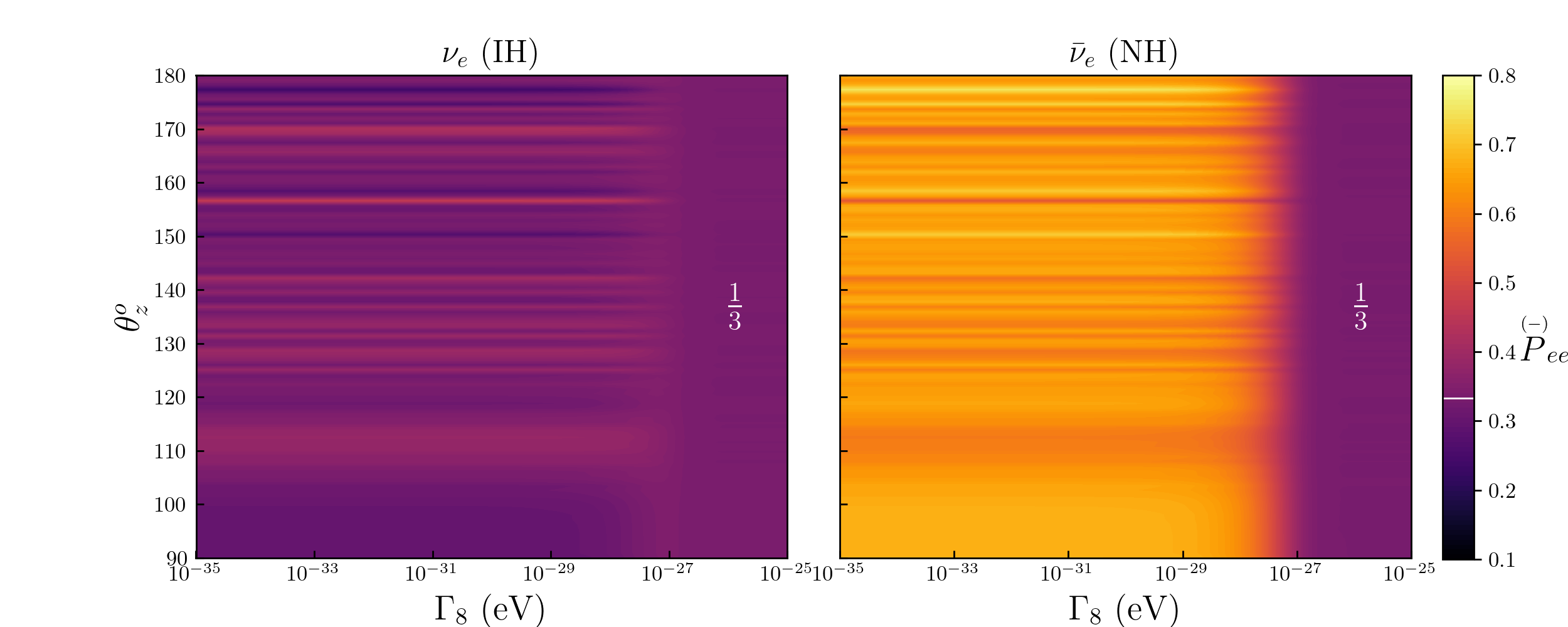}
    \caption{$P_{ee}$ in IH (left) and $\bar{P}_{ee}$ in NH (right) under Earth matter effects as a function of QD parameter for $E_\nu = 30$ MeV, considering a SN 10 kpc away from Earth and $n=0$. It is possible to see that QD suppresses regeneration effects for $\Gamma_8 \gtrsim 10^{-27}$ eV, where $\Gamma_3 = 10^{-32}$ eV was set. The white line on the color bar represents maximal mixing.}
\end{figure}

In Fig.~\ref{fig:regeneration-g} we also see the QD effects (MSC$^\epsilon$ with $n=0$) combined with Earth matter effects for a specific energy (similarly as shown in Fig.~\ref{fig:pee-combined}, but for a wide range of $\theta_z$ and the QD parameter). The asymptotic maximal mixing suppresses regeneration effects beyond $\Gamma_8 \sim 10^{-27}$ eV, being a leading effect. Since regeneration is a second-order effect, we impose bounds on QD in the next section without considering Earth matter effects, and by the end of Section~\ref{sec:msc}, we show its impact on results.

\section{Future limits on quantum decoherence}\label{sec:future-limits}

In order to impose bounds on QD using simulated data, we perform a binned $\chi^2$ through pull method \cite{fogli2002getting} over QD parameters for MSC and $\nu$-loss scenarios:

\begin{equation}\label{eq:chi}
    \chi^2 = \sum_{d} \sum_{j=1}^m \frac{(N_{j,d}^\text{true} - (1+a) N_{j,d}^\text{th})^2}{N_{j,d}^\text{th}} + \frac{a^2}{\sigma_a^2}
\end{equation}
where $m$ indicates the number of energy bins, $d$ represents each detector, $N_{j,d}^\text{true}$ represents events predicted by the MSW solution, and $N_{j,d}^\text{th}$ accounts the theoretical number of events of the marginalized model in our analysis, i.e. MSW + quantum decoherence  respectively and the second term on the right-hand side takes our estimation in the flux uncertainties of $40\%$ into account \cite{de2020impact}.

We can note in Fig.~\ref{fig:Pab_loss} that since all probabilities vanish for high values of $\gamma$, $N \rightarrow 0$ for $\nu$-loss. However in order to avoid a bias in our analysis, we marginalize over $\gamma$ only in a range where the requirement of at least $\sim 5$ events per bin is achieved (we use the same rule for MSC). We also take the size of the bins to be twice the detector energy resolution. Using these requirements, JUNO allows a single bin for $\nu$-loss, being a counting experiment for this analysis. The bins scheme for DUNE and HK are also changed for $\nu$-loss compared to MSC in order to match the established minimum number of events per bin in the tested range of $\gamma$.

\begin{figure}
    \centering
    \includegraphics[width=0.9\textwidth]{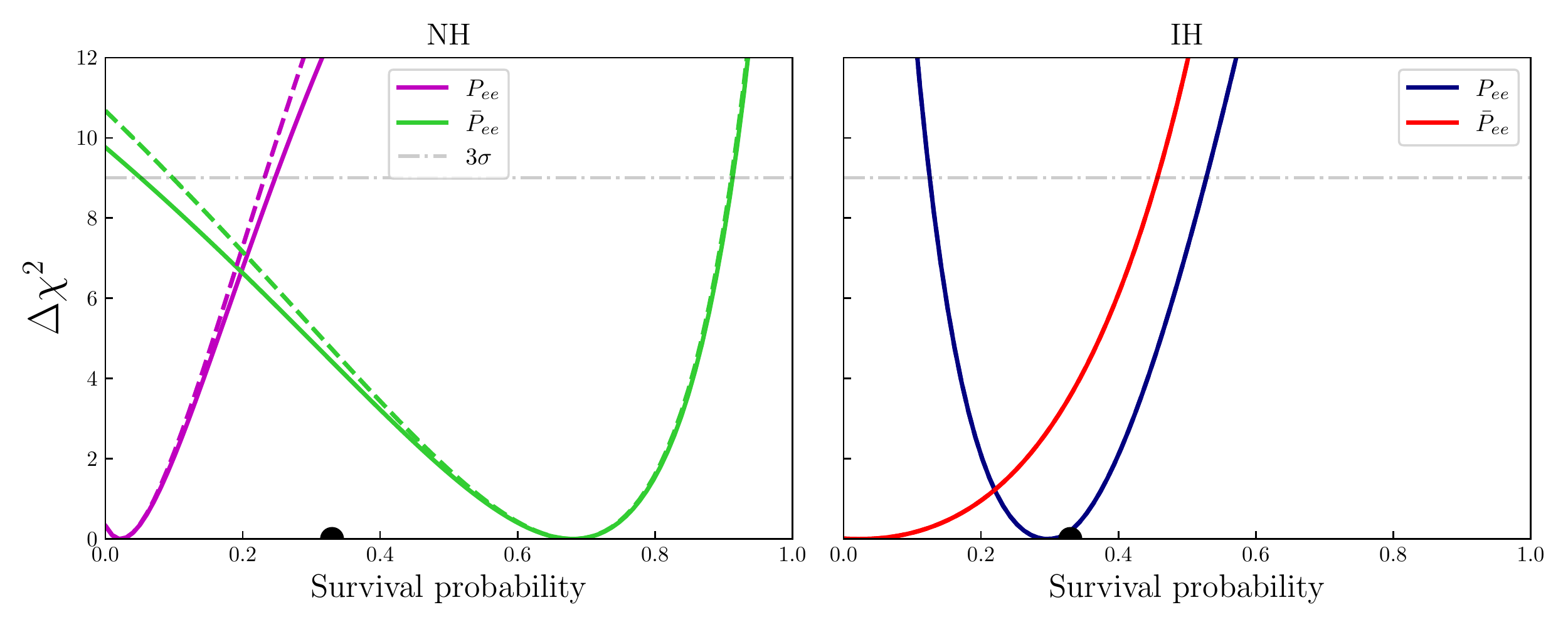}
    \caption{Limits on $P_{ee}$ and $\bar{P}_{ee}$ for the 27 $M_\odot$ (solid) and 11.2 $M_\odot$ (dashed) progenitor stars from simulations, considering only the neutronization burst. No quantum decoherence effects are taken into account in this Figure. The distance from Earth considered was 10~kpc. The probability is assumed to be a free parameter as recently proposed in \cite{dedin2023sn1987a}. The assumption of a standard adiabatic MSW conversion at the SN is taken into account (as all along the manuscript), getting rid of the energy dependency on $P_{ee}$ and $\bar{P}_{ee}$. The black dot is the maximal mixing scenario (1/3). Note that the 11.2 $M_\odot$ line for IH matches to the 27 $M_\odot$, showing that the sensitivity for simulated progenitors tested is similar.}
    \label{fig:pee-profile}
\end{figure}

Before imposing limits on MSC and $\nu$-loss with eq.~(\ref{eq:chi}), we can treat $P_{ee}$ and $\bar{P}_{ee}$ as free parameters, which is a reasonable approximation to an adiabatic propagation at the SN, since these probabilities are energy independent (see \cite{dedin2023sn1987a} for a more detailed discussion in the context of SN1987A), we perform a marginalization with $\chi^2(P_{ee},\bar{P}_{ee})$ in eq.~(\ref{eq:chi}) to understand how far asymptotically QD scenarios are from the standard $\nu$ mixing and also see how sensible a combined measurement (DUNE+HK+JUNO) could be, using uniquely the neutronization burst. Fig.~\ref{fig:pee-profile} shows how a 10 kpc SN can impose limits to $P_{ee}$ and $\bar{P}_{ee}$, with NH and IH concerning the true MSW model. The black dot represents maximal mixing or the asymptotic limit of MSC, which is closer to the IH solution (given by the corresponding best-fit value) than NH for $P_{ee}$, but in an intermediary point of hierarchies with respect to $\bar{P}_{ee}$. In the $\nu$-loss scenario it is not so clear from Fig.~\ref{fig:pee-profile} which hierarchy would lead to stronger constraints, given the presence of other probabilities, such as the ones in Fig.~\ref{fig:Pab_loss}. 

Using eq.~(\ref{eq:chi}) and the procedures described in Sections \ref{sec:qd-in-sn} and \ref{sec:metho-sim}, we treat QD parameters as free and perform a $\chi^2$ analysis in order to impose statistical bounds in this effect using a future SN detection. Since nowadays the neutrino mass hierarchy is not established, we include both scenarios in our analysis. 

We test both MSW-NH versus the marginalized MSW-NH + QD and also the MSW-IH versus the marginalized MSW-IH + QD in order to understand how restrictive future detectors will be. The results will show that if QD plays any role in SN neutrinos, both possible $\nu$ hierarchies could be affected.

\subsection{MSC$^\slashed{\epsilon}$}\label{sec:msc-conserved}

\begin{figure}[h]\label{fig:limits-mass-state-all-detec-cons-E}
    \centering
    \includegraphics[width=\textwidth]{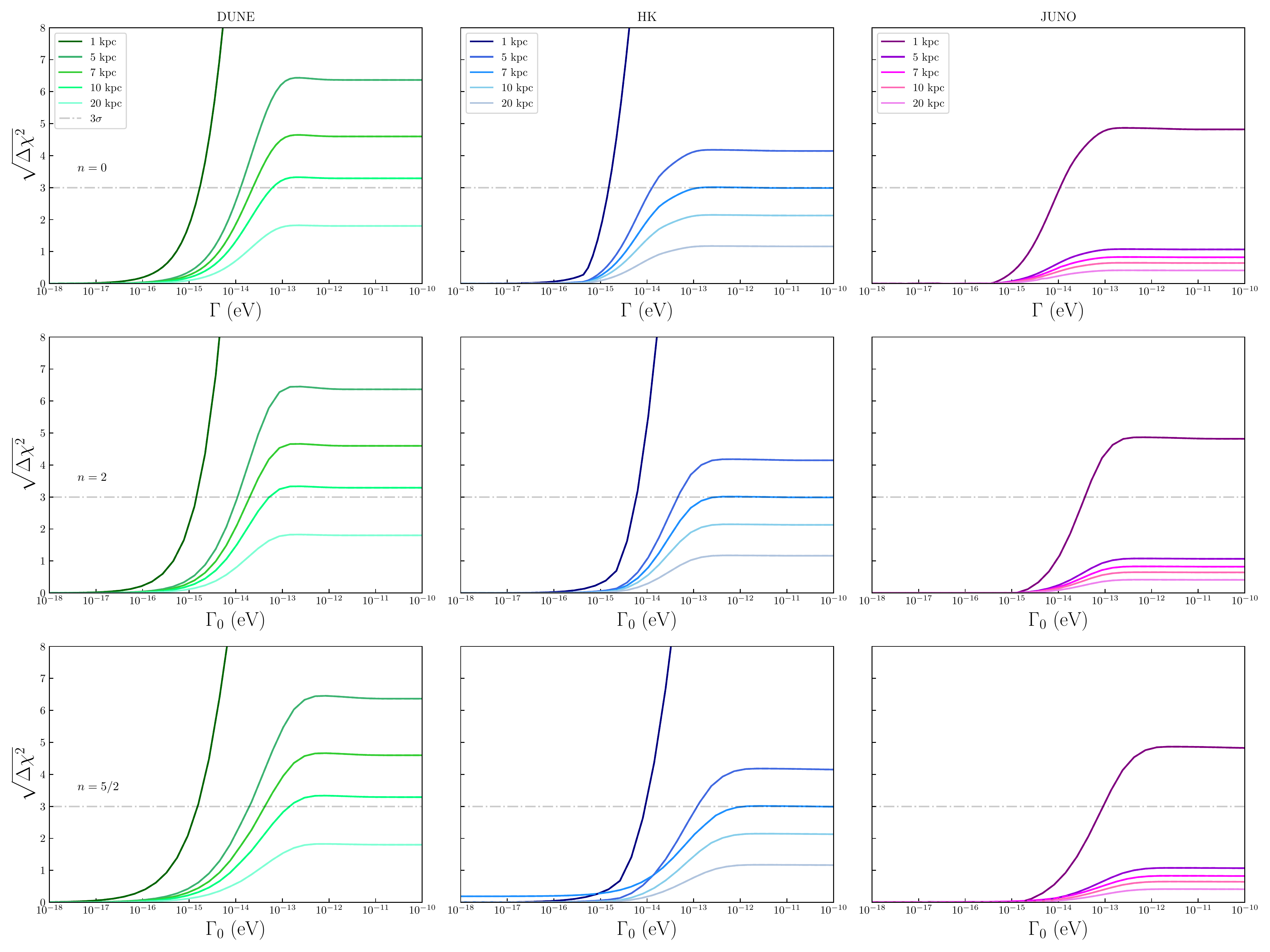}
    \caption{Limits on $\Gamma$ for various SN distances from Earth for DUNE (left), HK (middle), and JUNO (right) for the 40~$M_\odot$ progenitor star simulation. The true scenario taken into account was NH, and we marginalize the parameters over the theoretical NH+QD (MSC$^{\slashed{\epsilon}}$). No Earth matter effect was considered. Each row means a different value of $n$ in the parameterization $\Gamma = \Gamma_{0} (E/E_0)^n$.}
\end{figure}

For the MSC$^\slashed{\epsilon}$ model, we calculate the $\sqrt{\Delta\chi^2}$ bounds over the parameter $\Gamma$, where $\Delta\chi^2 = \chi^2 - \chi^2_\text{min}$ (since we are not including statistical and systematic uncertainties when producing the ``true" data, we always have $\chi^2_\text{min}=0$).
The results for the 3 experiments are summarized in Fig.~\ref{fig:limits-mass-state-all-detec-cons-E}, where the true scenario is NH and we marginalize over NH+QD. Note that bounds reach different significant limits for each SN distance, with lower distances being more restrictive.

Since the traveled distance is a fixed feature, the only aspect that the SN distance from Earth contributes is the number of events detected. Following Fig.~\ref{fig:limits-mass-state-all-detec-cons-E}, the best performance in NH is for DUNE, with possible $3\sigma$ limits for a 10 kpc  SN away from Earth of:

\begin{equation}
\Gamma_0 \leq \left\{
\begin{array}{rl}
6.2 \times 10^{-14}  ~{\rm eV} ~~~&(n=0)  \\
5.2 \times 10^{-14} ~{\rm eV} ~~~&(n=2)   \\
1.4 \times 10^{-13} ~{\rm eV} ~~~&(n=5/2)     
\end{array}
\right.
\end{equation}

For a SN at a distance of 1 kpc, limits of $\mathcal{O}(10^{-16})$~eV can be reached.
HK has also a good performance and achieves $2\sigma$ bounds for a 10 kpc SN. JUNO is not capable of individually achieving reasonable bounds on QD for SN distances $\gtrsim 1$ kpc, but would also have a strong signal for a galactic SN as close as 1 kpc away from Earth, which can be attributed to the small fiducial mass compared to HK and a single IBD channel considered in this work (with a significantly lower cross-section than $\nu_e$-Ar for energies above $\sim 15$ MeV). Other channels, such as $\nu$-p elastic scattering could possibly improve the results, but given the detection challenges associated, we decided to not include them here.

\begin{figure}[h]\label{fig:limits-mass-state-all-detec-cons-E-ih}
    \centering
    \includegraphics[width=\textwidth]{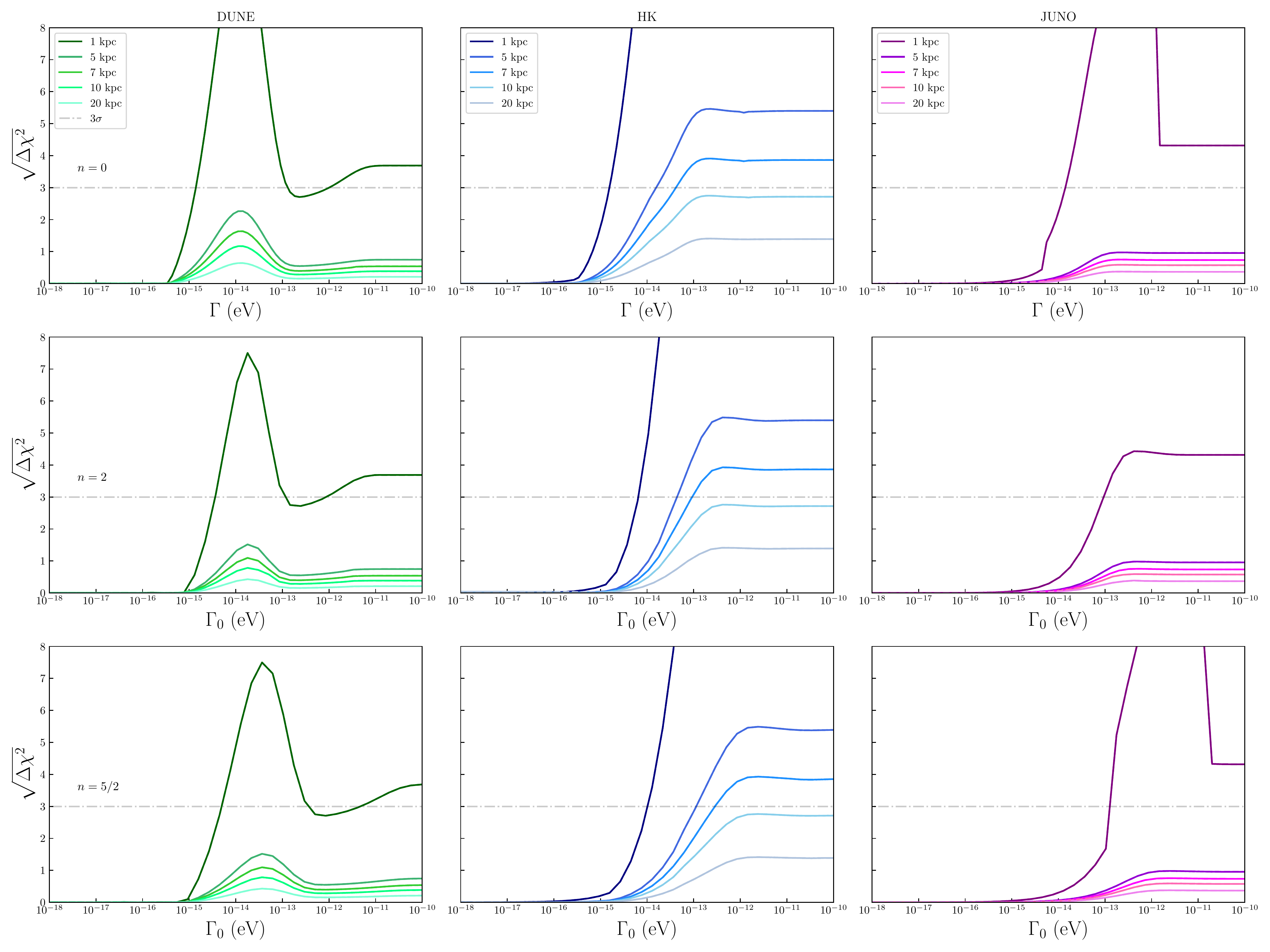}
    \caption{Same as Fig.~\ref{fig:limits-mass-state-all-detec-cons-E} but with IH as the true theory, marginalized over the parameters of the IH+QD model. }
\end{figure}

We also performed the same analysis using IH as the true theory and marginalizing over IH+QD. The results are shown in Fig.~\ref{fig:limits-mass-state-all-detec-cons-E-ih}. The best performance is clearly for HK with $2\sigma$ bound of:

\begin{equation}
\Gamma_0 \leq \left\{
\begin{array}{rl}
3.6 \times 10^{-14} ~{\rm eV} &~~~(n=0) \\
8.0 \times 10^{-14} ~{\rm eV} &~~~(n=2)  \\
2.4 \times 10^{-13} ~{\rm eV} &~~~(n=5/2)  
\end{array} 
\right.
\end{equation}
for a 10 kpc SN from Earth. DUNE is not capable to impose strong bounds in an IH scenario. JUNO performance is improved for distances $\lesssim 1$ kpc compared to NH. Results are summarized in Table~\ref{tab:results-msc-noe} in Appendix~\ref{appendix:b}.

A 20 kpc SN could not impose strong bounds for individual experiments. Distances as far as 50~kpc (as Large Magellanic Cloud) were not investigated in this work, given the lack of events per bin, in which a more refined unbinned statistical analysis would be required, which is not strongly motivated by the fact that expected limits are below $2\sigma$.

The bounds and sensitivity of each detector in a given hierarchy shown above could be associated with the sensitivity to ${P}_{ee}$ and $\bar{P}_{ee}$ shown in Fig.~\ref{fig:pee-profile}. In NH (left plot), limits over $P_{ee}$ are more restrictive than $\bar{P}_{ee}$ with respect to maximal mixing represented by the black dot. For IH (right plot), we have an opposite sensitivity, since $P_{ee} \sim 1/3$, while for $\bar{P}_{ee}$ there is a gap between the best fit and 1/3 probability, allowing limits with certain significance to be imposed. Since DUNE is most sensitive to $\nu_e$, via $\nu_e$-Ar interaction, it will be more sensitive to $P_{ee}$ and then more relevant in the NH scenario. As for HK and JUNO, they are more sensitive to $\bar\nu_e$ and therefore to $\bar{P}_{ee}$, which reflects a better performance in the IH scenario. In our calculations, the elastic scattering considered in HK does not contribute much to the total $\chi^2$.

\subsection{MSC$^\epsilon$}\label{sec:msc}

\begin{figure}[h]\label{fig:limits-mass-state-all-detec}
    \centering
    \includegraphics[width=\textwidth]{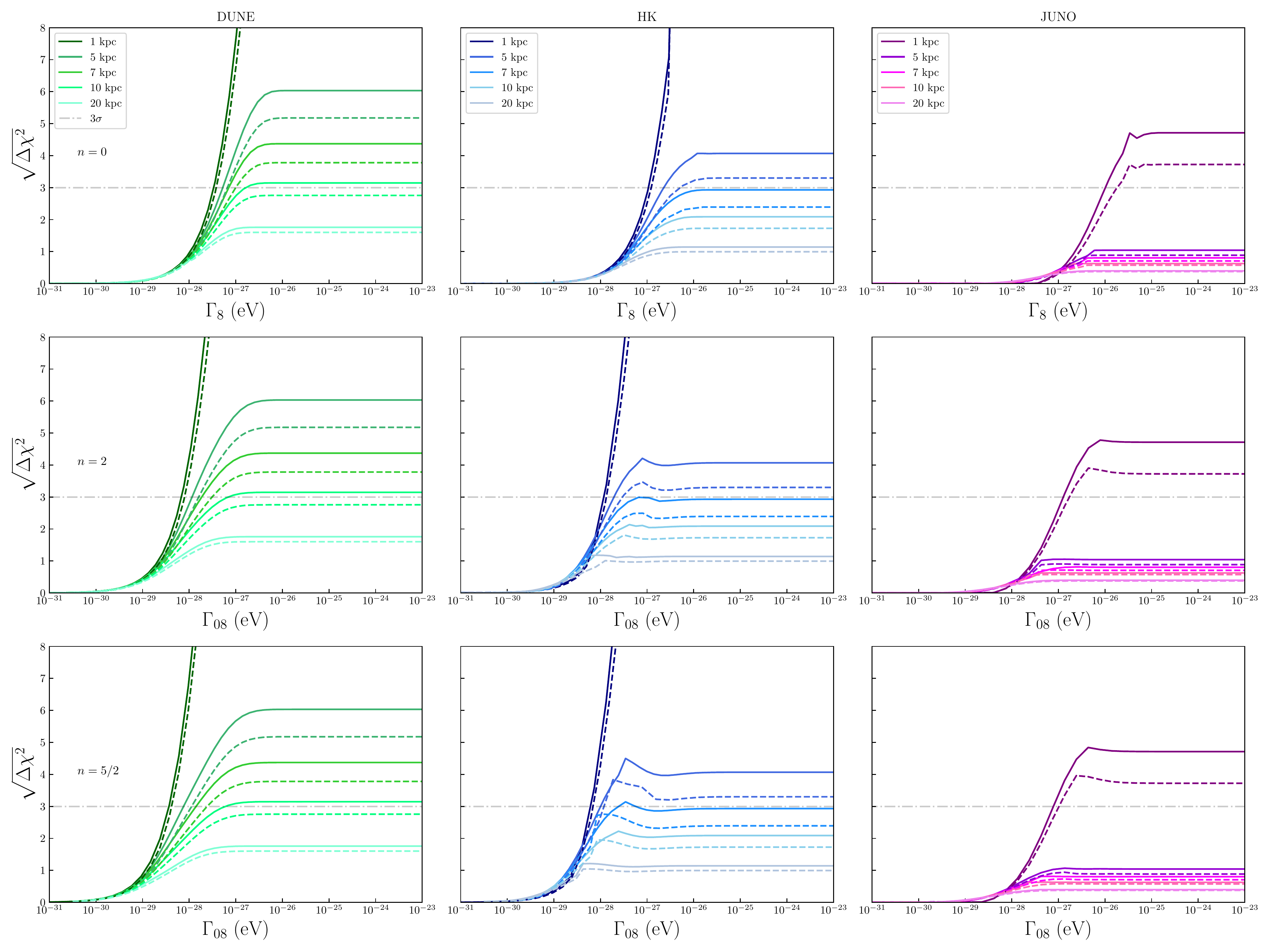}
    \caption{Same as \ref{fig:limits-mass-state-all-detec-cons-E} but for MSC$^\epsilon$ with simulations of the 27 $M_\odot$ (solid) and 11.2 $M_\odot$ (dashed) progenitor masses. The bounds are orders of magnitude more restrictive than for MSC$^{\slashed{\epsilon}}$.}
\end{figure}

The same procedure described in the section above was performed on the MSC$^\epsilon$ model, with bounds over the parameter $\Gamma_8$. Results are summarized in Fig.~\ref{fig:limits-mass-state-all-detec} for NH vs NH+QD. SN distance also plays an important role in this scenario and results and their aspects are similar to MSC$^\slashed{\epsilon}$ described in the last section. DUNE has the best performance for the tested SN distances and even for a 10~kpc SN, bounds with $3\sigma$ could be achieved for $n = 0,2$ and $5/2$. 
Despite the stronger effects caused by MSC for larger distances, the number of events decrease with $L^2$, and stronger limits can be imposed for a SN happening at shorter distances, reflecting that the larger number of neutrinos arriving at the detector is a crucial aspect. 

\begin{figure}[h!]
    \centering
    \includegraphics[width=0.6\textwidth]{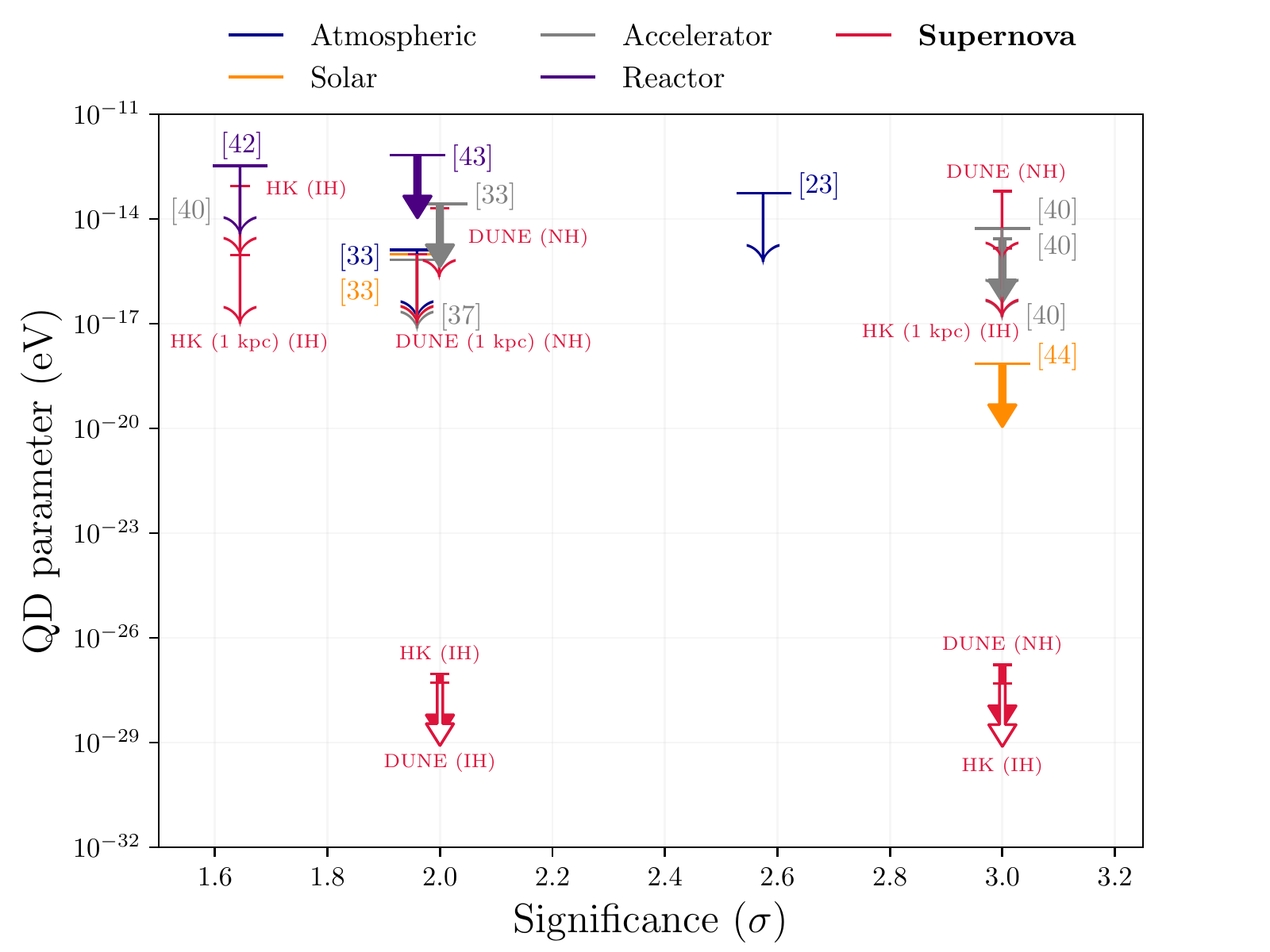}
    \caption{Current bounds on quantum decoherence for a number of works from many neutrino sources and also the SN limits presented here ($n=0$). Arrows with longer horizontal bases correspond to current experimental bounds, whereas minor bases are to respect to possible future limits. Numbers in the arrows indicate the reference in which the limits were obtained. Thin arrows indicate bounds equivalent to MSC$^\slashed{\epsilon}$, while thick-filled ones are the MSC$^{\epsilon}$. White-filled thick arrows correspond to $\nu$-loss bounds. Supernova limits described in this work are in red, and are to respect to a distance of 10~kpc from Earth unless distance is indicated, with more restrictive bounds being possible for closer SNs.}
    \label{fig:bounds}
\end{figure}

From Fig.~\ref{fig:limits-mass-state-all-detec}, taking the result of a 10 kpc SN (27 $M_\odot$), DUNE would potentially impose $\Gamma_8 \leq 4.2 \times 10^{-28}$ eV for $2\sigma$ and $\Gamma_8 \leq 1.7 \times 10^{-27}$ eV for $3\sigma$ with $n = 0$, whereas the HK bound is $\Gamma_8 \leq 4.2 \times 10^{-27}$ eV for $2\sigma$. Looking at limits from various works \cite{lisi2000probing,farzan2008reconciling, oliveira2014quantum, oliveira2016dissipative, coelho2017nonmaximal, coelho2017decoherence, carrasco2019probing, carpio2019testing,gomes2017parameter,coloma2018decoherence, de2020solar, gomes2019quantum}, to the best knowledge of the authors, this is an unprecedented level of sensitivity for testing quantum decoherence, orders of magnitude more restrictive than any other work in the subject.
Fig.~\ref{fig:bounds} shows bounds from works with different sources and place the limits from this work for both hierarchy scenarios. 

Note that for $n = 2$ and 5/2 the bounds are over $\Gamma_{08}$ in $\Gamma_8 = \Gamma_{08} (E/10~{\rm MeV})^n$. For a 10 kpc SN (27$M_\odot$), DUNE $3\sigma$ bounds reach:

\begin{equation}
\Gamma_{08} \leq \left\{
\begin{array}{rl}
7.0 \times 10^{-28} ~{\rm eV} &~~~(n=2)  \\ 
6.2 \times 10^{-28} ~{\rm eV} &~~~(n=5/2) 
\end{array} .
\right.
\end{equation}
HK is able to achieve $2\sigma$ bounds as restrictive as $\Gamma_{08} \leq 2.7 \times 10^{-28}$~eV and $\Gamma_{08} \leq 1.2 \times 10^{-28}$~eV for $n=2$ and 5/2 respectively. All mentioned results are summarized in Table~\ref{tab:results} in the Appendix~\ref{appendix:b}.

\begin{figure}[h]\label{fig:combined-mass-state}
    \centering
    \includegraphics[width=\textwidth]{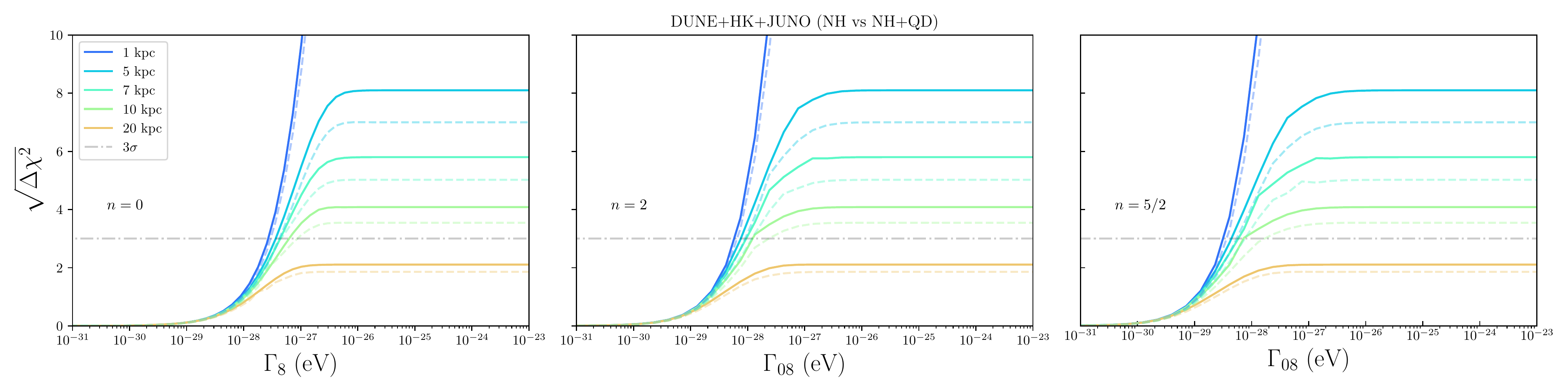}
    \caption{Combined fit for the true MSW-NH marginalizing over MSW-NH with QD (MSC$^\epsilon$) effects.}
\end{figure}

We also performed a combined fit for the three detectors using the same $\nu$ hierarchy scheme shown in Fig.~\ref{fig:combined-mass-state}, where a $3\sigma$ limit for a 10 kpc SN would reach:

\begin{equation}
\Gamma_{08} \leq 
\left\{
\begin{array}{rl}
6.2 \times 10^{-28} ~{\rm eV}     &  ~~~(n=0)\\
1.2 \times 10^{-28} ~{\rm eV}     &  ~~~(n=2)\\
0.72 \times 10^{-28} ~{\rm eV}     &  ~~~(n=5/2)
\end{array}     .
\right.
\end{equation}
Even a $4\sigma$ of maximal mixing is possible to be achieved for all values of $n$, but such significance is achieved only by the 27 $M_\odot$ simulated progenitor. Although a combined analysis reaches high significance, it should be taken with a grain of salt, since it is not possible to be sure that experiments would be simultaneously in operation.

\begin{figure}[h]\label{fig:limits-mass-state-coupling-ih-vs-ih-oqs}
    \centering    
    \includegraphics[width=\textwidth]{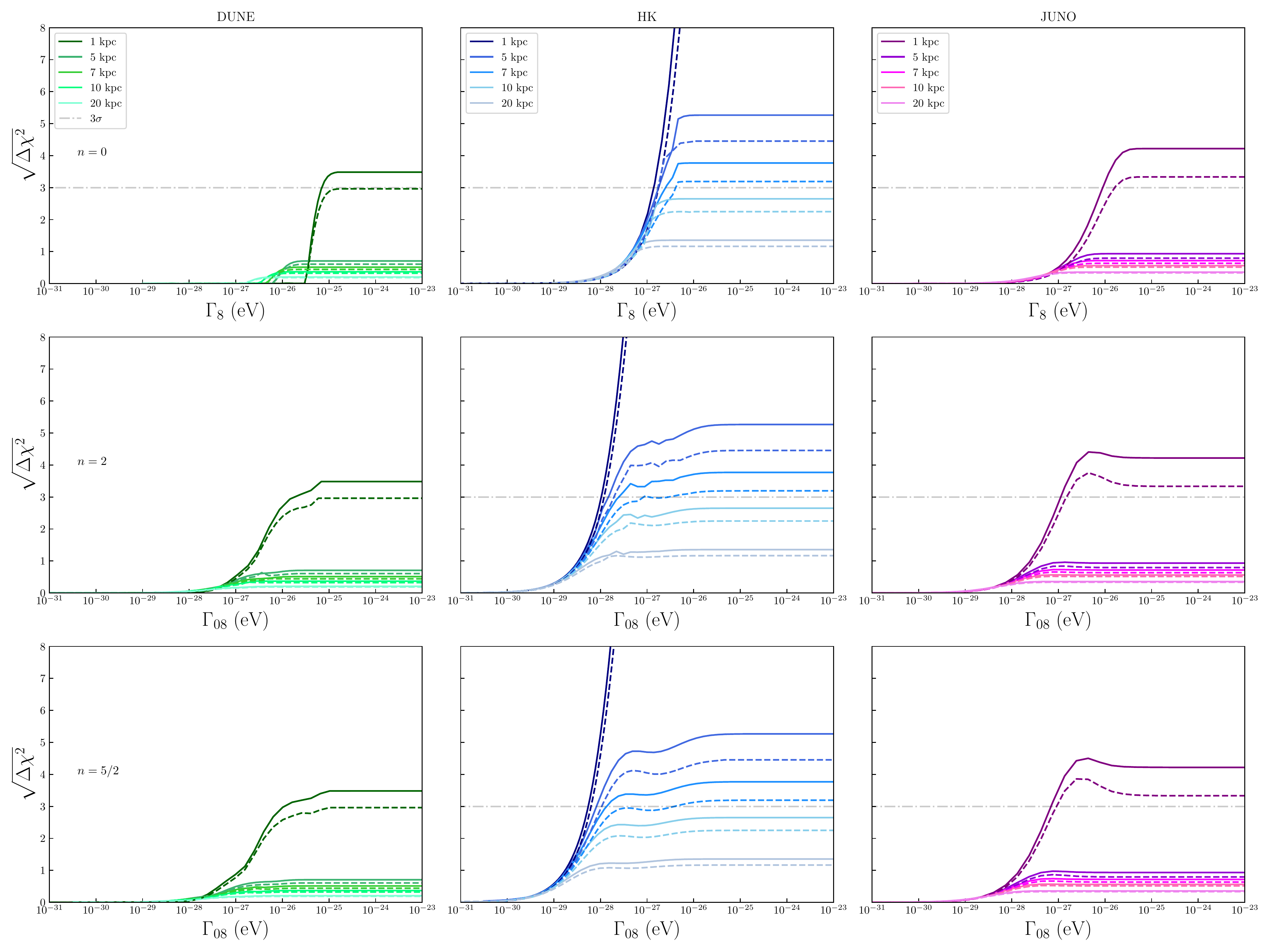}
    \caption{Same as Fig.~\ref{fig:limits-mass-state-all-detec} but for IH versus IH + QD.}
\end{figure}

\begin{figure}[h]\label{fig:combined-mass-state-ih-vs-ih-oqs}
    \centering
    \includegraphics[width=\textwidth]{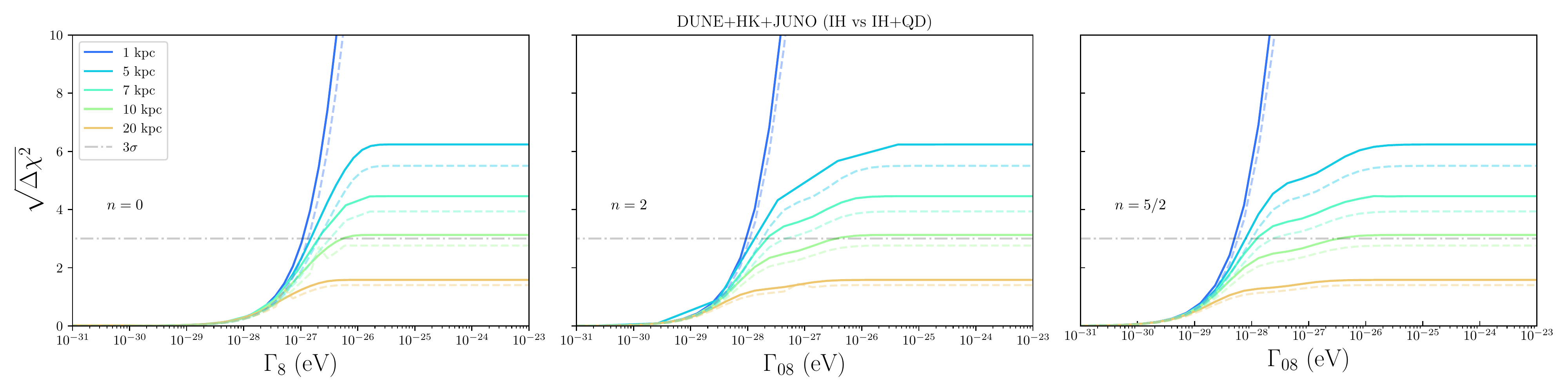}
    \caption{Same as Fig.~\ref{fig:combined-mass-state} but now accounting the IH scenario.}
\end{figure}

Using the same procedure as done in NH, we make the analysis assuming IH as the true mixing and marginalizing over IH+QD. The results are shown in Fig.~\ref{fig:limits-mass-state-coupling-ih-vs-ih-oqs}. HK has the strongest bounds on this scenario but does not reach $3\sigma$ for a 10 kpc SN, even though the potential limits for $2\sigma$ are:
\begin{equation}
\Gamma_{08} \lesssim \left\{
\begin{array}{rl}
 1.3 \times 10^{-27}  ~{\rm eV}     &  ~~~(n=0)\\
 1.4 \times 10^{-28}  ~{\rm eV}     &  ~~~(n=2)\\
 4.9 \times 10^{-28}  ~{\rm eV}     &  ~~~(n=5/2)\\
\end{array}    .
\right.
\end{equation}
DUNE has a very poor performance in this scenario for any distance $\gtrsim 1$ kpc. JUNO sensitivity is similar to NH marginalization discussed above. In a combined fit in IH, shown in Fig.~\ref{fig:combined-mass-state-ih-vs-ih-oqs}, the following $3\sigma$ limits can be obtained:
\begin{equation}
\Gamma_{08} \lesssim \left\{
\begin{array}{rl}
 5.4 \times 10^{-27}  ~{\rm eV}     &  ~~~(n=0)\\
 3.5 \times 10^{-27}  ~{\rm eV}     &  ~~~(n=2)\\
 3.3 \times 10^{-27}  ~{\rm eV}     &  ~~~(n=5/2)\\
\end{array}    .
\right.
\end{equation}

\begin{figure}[h]\label{fig:limits-regeneration}
    \centering
    \includegraphics[width=0.8\textwidth]{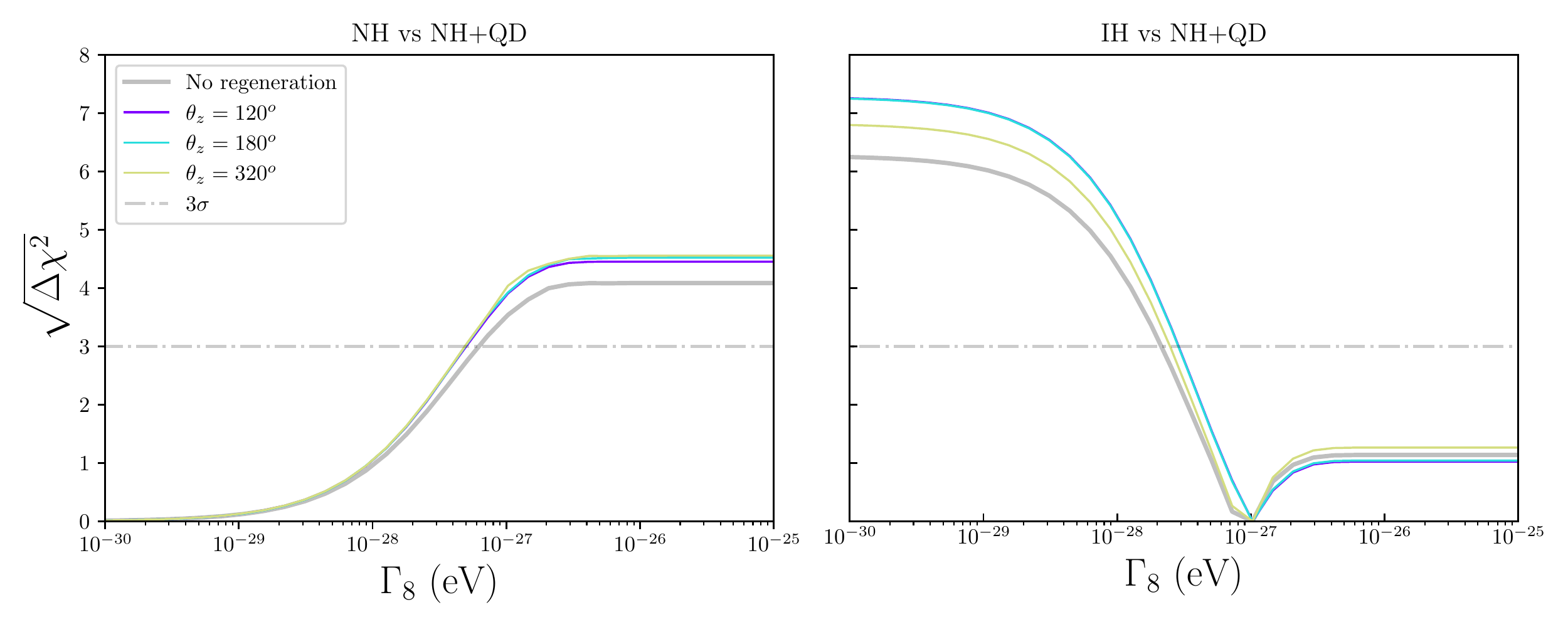}
    \caption{Limits on MSC with the impact of Earth matter effects for a SN 10 kpc from Earth and the 27~$M_\odot$ simulation for different zenith angles $\theta_z$ ($n=0$). The limits correspond to a combined detection of DUNE, HK, and JUNO, but $\theta_z$ is to respect to DUNE, with SN beam in the direction of DUNE longitude. The $\theta_z = 320^o$ means that regeneration effects at HK and JUNO are expected, even if the SN beam does not cross Earth for reaching DUNE.}
\end{figure}

To check the impact of regeneration on the above results, we calculated the bounds of a combined detection of DUNE, HK, and JUNO including this effect. We test different $\theta_z$, the zenith to respect to DUNE, with the assumption that the SN flux comes from DUNE longitude. The results are in Fig.~\ref{fig:limits-regeneration}. We can note in the left plot that the impact of the Earth matter effect is small but enhances QD bounds for a 10~kpc detection and limits could be stressed beyond $4\sigma$. The right plot shows the situation where the IH scenario is assumed to be true and NH+QD is marginalized. We will discuss such a scenario in Section~\ref{sec:nu-hierar-measur}, but we also see that regeneration will not change significantly the results.

\subsection{Neutrino loss}

\begin{figure}[h]\label{fig:limits-nu-loss}
    \centering
    \includegraphics[width=\textwidth]{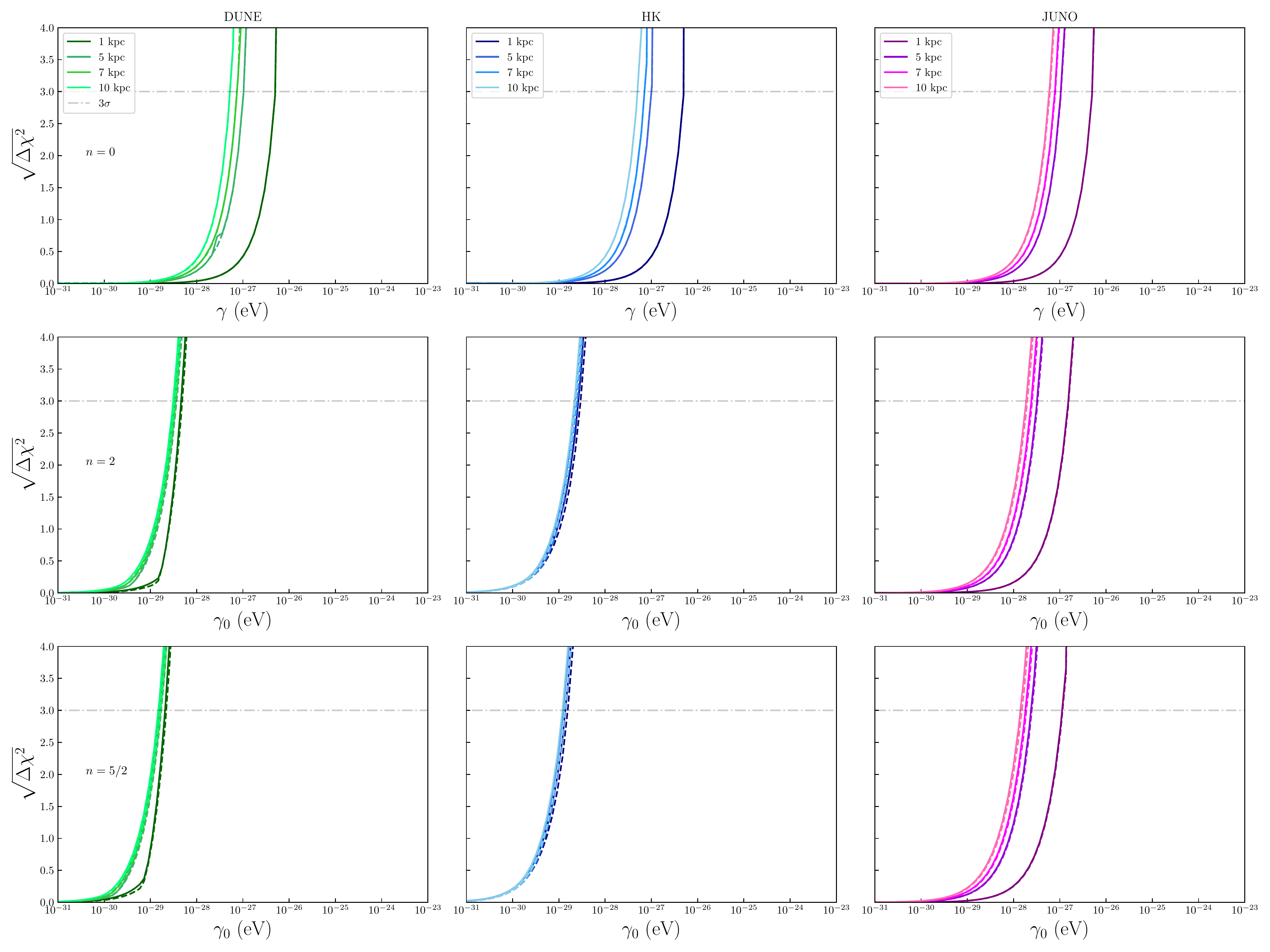}
    \caption{Limits on $\gamma$ for various SN distances from Earth for all detectors in the $\nu$-loss scenario with true IH marginalized over the parameters of the IH+QD model.}
\end{figure}

Since in $\nu$-loss the spectrum of events decreases asymptotically to zero, the bounds on this scenario are expected to be as significant or even more than MSC for all experiments. Since the calculated number of events for NH is low (mainly for DUNE and JUNO) and $\nu$-loss would decrease it, not fulfilling our requirement of $\gtrsim 5$ events per bin, we perform here only the IH (true) versus IH+QD. Fig.~\ref{fig:limits-nu-loss} shows the $\sqrt{\Delta\chi^2}$ for each individual detector. We see that high values of $\gamma$ are strongly bounded, even for JUNO. For a SN from 10~kpc away from Earth, DUNE, HK and JUNO are capable to impose $\gamma \leq 5.2 \times 10^{-28}$~eV, $\gamma \leq 4.9 \times 10^{-28}$~eV and $\gamma \leq 5.9 \times 10^{-28}$~eV respectively with $3\sigma$ of significance ($n=0$). Note that beyond 10~kpc the number of events per bin would be significantly small for a $\nu$-loss scenario and we do not consider it in this analysis.

HK is capable to achieve the best ($3\sigma$) bounds with $\gamma_0 \leq 2.1 \times 10^{-29}$~eV and $\gamma_0 \leq 1.2 \times 10^{-29}$~eV for $n=2$ and $5/2$ respectively, with a 10 kpc SN. Although not shown in the plots, it is worth mentioning that HK would impose bounds on $\gamma$ even for NH, given the high statistics associated with this experiment, being the most sensitive one for the $\nu$-loss model. We detail the bounds and all mentioned results here in Table~\ref{tab:results-loss}.

\section{Neutrino mass hierarchy measurement}\label{sec:nu-hierar-measur}

\begin{figure}[h]
    \centering
    \includegraphics[width=\textwidth]{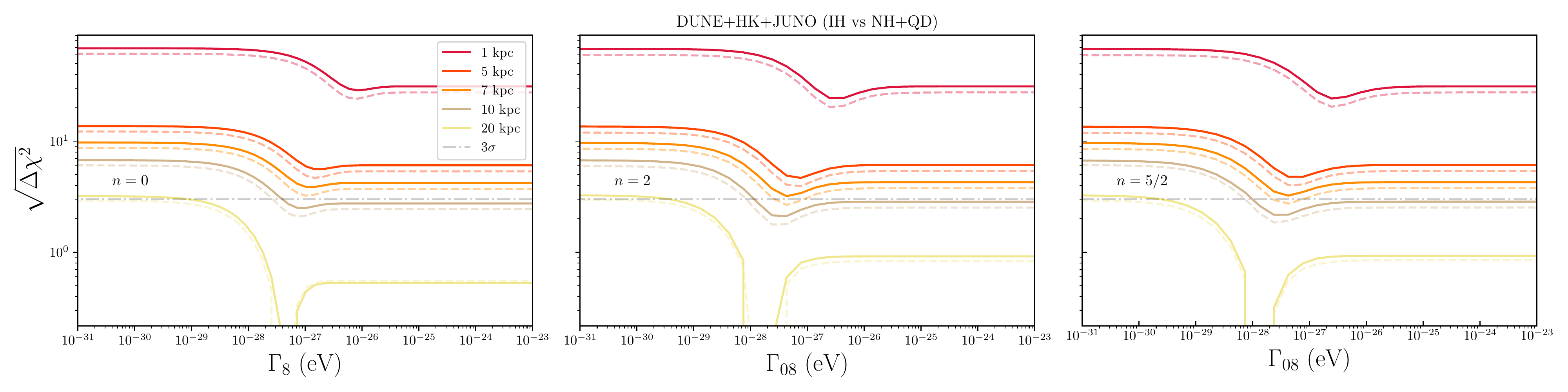}
    \caption{Statistically comparing the inverted hierarchy (IH) to normal hierarchy (NH) with the impact of quantum decoherence for a combined detection using the 11.2~$M_\odot$ (dashed) and 27~$M_\odot$ (solid) simulations. No regeneration effects were taken into account.}
    \label{fig:mass_state_ih_vs_nhqd}
\end{figure}

In a future supernova detection, the neutronization burst arises as a robust test of neutrino mass hierarchy, with $\nu$-Ar in DUNE capable to determine the correct  scenario with relatively high confidence.
However, although the possible strong bounds are to be imposed on quantum decoherence, if QD plays a significant role in $\nu$ mixing, the IH could be mimicked by a NH with the impact of QD (particularly, in the MSC models). A similar analysis was performed in the context of $\nu$-decay in \cite{delgado2022probing}. Therefore, the question that arises is how much NH and IH are distinguishable if we compare both hierarchies superposing the standard NH to QD. Fig.~\ref{fig:mass_state_ih_vs_nhqd} shows the statistical bounds of the scenario where IH is taken as the true theory and NH+QD is marginalized in a combined detection for $n=0,2,5/2$. The results show that the significance of hierarchy determination significantly weakens for the tested SN distances and even a combined detection could not disentangle the hierarchies if MSC plays an important role. 

To check this statement we can compare the values of $\sqrt{\Delta\chi^2}$ for $\Gamma_8 \rightarrow 0$ and $\Gamma_8 \rightarrow \infty$ in Fig.~\ref{fig:mass_state_ih_vs_nhqd}. We can assume that $\sqrt{\Delta\chi^2}|_{\Gamma_8 \rightarrow 0}$ corresponds to the distinguishability of hierarchy in a standard scenario
since $\Gamma_8$ is small enough to neglect QD effects. The plateau in the limit of $\sqrt{\Delta\chi^2}|_{\Gamma_8 \rightarrow \infty}$ shows how NH+QD would differ from IH in a future combined detection, in which has lower values of $\sqrt{\Delta\chi^2}$, resulting in a less significant hierarchy discrimination. Taking as a reference a SN distance of 10~kpc for the 27~$M_\odot$ simulation, with a combined detection of DUNE, HK and JUNO, we have a $\sqrt{\Delta\chi^2}|_{\Gamma_8 \rightarrow 0} = 6.89$ going to $\sqrt{\Delta\chi^2}|_{\Gamma_8 \rightarrow \infty} = 3.13$. For an individual detection with the same SN distance, DUNE would change from $\sqrt{\Delta\chi^2}|_{\Gamma_8 \rightarrow 0} = 5.70$, which is statistically significant to determine the hierarchy, to a mere $\sqrt{\Delta\chi^2}|_{\Gamma_8 \rightarrow \infty} = 0.37$. HK also could be affect with a $\sqrt{\Delta\chi^2}|_{\Gamma_8 \rightarrow 0} = 3.36$ going to $\sqrt{\Delta\chi^2}|_{\Gamma_8 \rightarrow \infty} = 2.65$. JUNO can not distinguish the neutrino hierarchies significantly at 10~kpc. It is important to mention that for 1~kpc and 5~kpc DUNE could be highly affected by this hierarchy misidentification, but HK still would provide a distinction of $\gtrsim 5 \sigma$ even with QD effects. For SN distances $> 5$ kpc, the neutrino hierarchies would be hardly disentangled by the tested experiments if QD effects are significant. As far as we tested, the $\nu$-loss model did not lead to the same potential hierarchy misidentification found in the MSC.

\section{Conclusions}
\label{sec:conclusions}

In this paper, we have explored the capability of a future SN neutrino detection in imposing limits in quantum decoherence scenarios. As the neutrinos are already treated as an incoherent mixture of mass eigenstates inside the SN, damping effects are not expected, then we explore secondary quantum decoherence scenarios, such as the relaxation mechanism, which can be potentially observed in a SN neutrino signal. We limit ourselves to scenarios where the decoherence matrix $D$ is diagonal in the neutrino vacuum mass basis. Among the possible models to be investigated, we consider the ones we denoted as Mass State Coupling (MSC), leading to maximal mixing of states, and the neutrino loss ($\nu$-loss), associated to the loss of neutrino flux along propagation. These scenarios are well-motivated by quantum gravity, where a possible dependency with energy is expected in the form of $\gamma = \gamma_0 (E/E_0)^n$, and therefore, we explore the limits on the decoherence parameters for different $n$.

The analysis was done considering DUNE, HK, and JUNO as possible detectors. For the neutrino flux data, three progenitor stars were considered, a 40~$M_\odot$ (LS180-s40.0), 27 $M_\odot$ (LS220s27.0c) and 11.2 $M_\odot$ (LS220s11.2c), using the SN simulation data from the Garching group \cite{Garching, serpico2012probing, mirizzi2016supernova}. To get around the unsolved problem of neutrino collective effects, only the neutronization burst was considered, given that collective effects are expected to not play a significant role in this emission phase.

When considering the neutrino propagation inside the supernova, the relaxation effect could affect the neutrino flavor conversion, even with the assumption of no exchange of neutrino energy to the environment, or $[H,V_p] = 0$ (MSC$^\slashed{\epsilon}$).  We show that in this regime it is possible to get competitive limits to QD parameters. However, the required values for the decoherence parameters need to be much larger than the ones in the scenario where $[H,V_p] \neq 0$ (MSC$^\epsilon$) (see the Appendix~\ref{appendix:a}), which would provide the most restrictive bounds on QD to date. For MSC$^\epsilon$, we only consider the decoherence/relaxation acting on neutrino propagation in the vacuum from the SN until it reaches the detectors at Earth, for which the propagation length is orders of magnitude larger than the SN size, and therefore, more sensible to the relaxation effects. We also explore the possible effects of Earth regeneration due to the neutrino propagation inside the Earth, which has minor effects in the bounds for the relaxation parameters, being the vacuum propagation the most relevant coherence length.

With all considerations, we show that the detectors used in the analysis are capable to impose the limits listed in Tables~\ref{tab:results-msc-noe} and \ref{tab:results} for the MSC scenario, depending on the distance being considered and the neutrino mass hierarchy. For the NH, the DUNE detector is the most promising one, while HK is the most sensible in the case of IH. The possible limits on the decoherence parameters are orders of magnitude stronger than the ones imposed by current terrestrial and solar experiments, as shown in Fig.~\ref{fig:bounds}. For the $\nu$-loss scenario, the limits are shown in Table~\ref{tab:results-loss}. Due to the neutrino disappearance, extra care needed to be taken in this scenario so that the requirement of at least 5 events per bin is fulfilled and the $\chi^2$ analysis can be applied.

Finally, we explored the possible degeneracy between the different standard scenarios of unknown mass hierarchy (NH and IH) without QD and the ones with QD effects included. As we saw, the IH scenario could be easily mimicked by NH combined with QD-MSC effects.

\section*{Acknowledgements}

We thank Hans-Thomas Janka and the Garching group for providing the SN simulations used in this work. MVS is thankful to Alberto Gago for pointing out complete positivity relations. EK is very grateful for the hospitality of GSSI during the preparation of this manuscript.
This study was financed by the Coordenação de Aperfeiçoamento de Pessoal de Nível Superior - Brasil (CAPES) - Finance Code 001, and partially by the Fundação de Amparo à Pesquisa do Estado de São Paulo (FAPESP) grants no. 2019/08956-2, no. 14/19164-6, and no. 2022/01568-0.

\bibliographystyle{kpmod}
\bibliography{references}

\vspace{30pt}

\appendix

\pagebreak
\FloatBarrier
\section{Decoherence inside the SN and matter effects}\label{appendix:a}

The neutrino Hamiltonian in flavor basis affected by the charged current potential $V_W$, i.e. $H_f = H_f^\text{vac} + V_W$, can be diagonalized to $H_m$ by a unitary transformation provided by $U_m$ as

\begin{equation}
    \rho_f = U_m \rho_m U_m^\dagger \hspace{1cm} H_f = U_m H_m U_m^\dag ,
\end{equation}
getting the most general form of (\ref{eq:gksl}) in the effective neutrino mass basis in matter

\begin{equation}\label{eq:gksl-matter}
    \frac{d\rho_m}{dt} = -i[H_m,\rho_m] - [U_m^\dag \Dot{U}_m, \rho_m] + \sum_p^{N^2-1} (V_{pm} \rho_m V_{pm} - \frac{1}{2} \{V_{pm}^2, \rho_{m}\})
\end{equation}
or following the notation in (\ref{eq:gksl-liouvillian})

\begin{equation}\label{eq:gksl-liouvillian-m}
    \ket{\dot{\rho}_m} = -2 \mathcal{L}_m(t) \ket{\rho_m}
\end{equation}
For all purposes of this work, the propagation is adiabatic, or $\dot U_m = 0$ in (\ref{eq:gksl-matter}).

\begin{figure}
    \centering
    \includegraphics[width=0.45\textwidth]{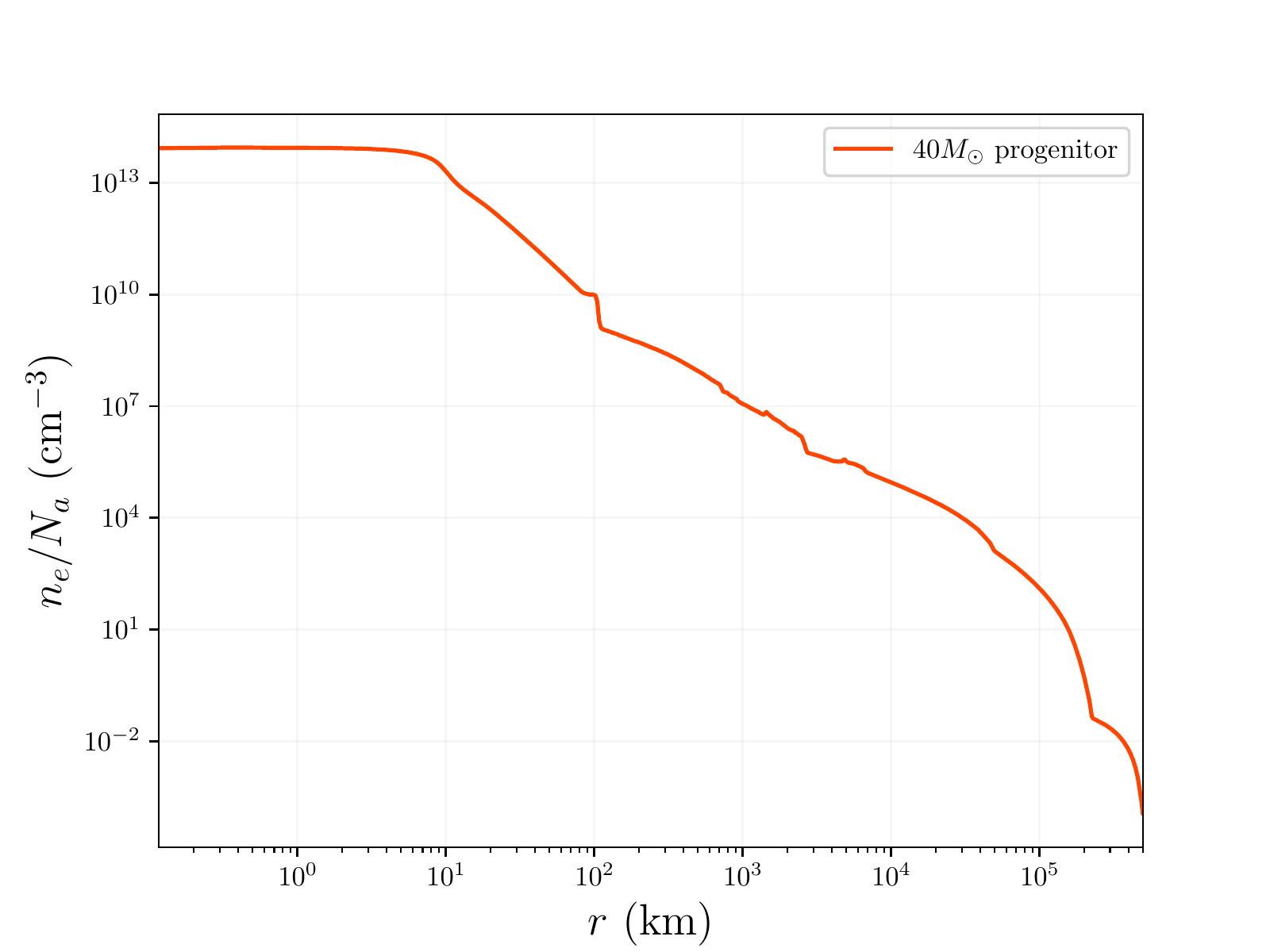}
    \caption{Snapshot (at 27~ms after the core bounce) of simulated SN electron density profile from the $40$~$M_\odot$ progenitor mentioned in the text \cite{serpico2012probing, Garching}.}
    \label{fig:density-sn}
\end{figure}

We are interested in solving equation~(\ref{eq:gksl-liouvillian-m}) in a variable matter density in order to get transition probabilities $P_{ij}^{m(\text{SN})}$ and $\bar{P}_{ij}^{m(\text{SN})}$. It is straightforward to obtain $\ket{\rho}$ in (\ref{eq:gksl}) , but in the case of $\ket{\rho_m}$, $V_{pm}$ and $H_m$ are time-dependent and the solution is a time-ordered exponential:

\begin{equation}\label{eq:time-ord}
    \mathcal{T}\left\{e^{-2\int^t_{t_0} dt^\prime \mathcal{L}_m(t^\prime)}\right\} = 1 + (-2)\int_{t_0}^t dt_1 \mathcal{L}_m(t_1) + (-2)^2\int_{t_0}^t dt_1 \int_{t_0}^{t_1} dt_2 \mathcal{L}_m(t_1) \mathcal{L}_m(t_2) + \cdots .
\end{equation}

Analytical solutions for specific cases in a variable matter density can be found in \cite{benatti2005dissipative, coloma2018decoherence}. However, instead of using a cumbersome approximated approach, we analyze the neutrino evolution into the SN making the limits in the integrals in (\ref{eq:time-ord}) $\Delta t = t_n - t_{n-1} \rightarrow 0$, allowing to solve (\ref{eq:time-ord}) numerically through slab approach, i.e. we divided the SN matter density profile into small parts, in which the neutrino Hamiltonian is approximately constant, then we make the time evolution from each step to another until the neutrino reach the vacuum. We use the simulated density profile in Fig.~\ref{fig:density-sn} to perform this calculation.

\begin{figure}[h!]
    \centering
    \includegraphics[width=0.4\textwidth]{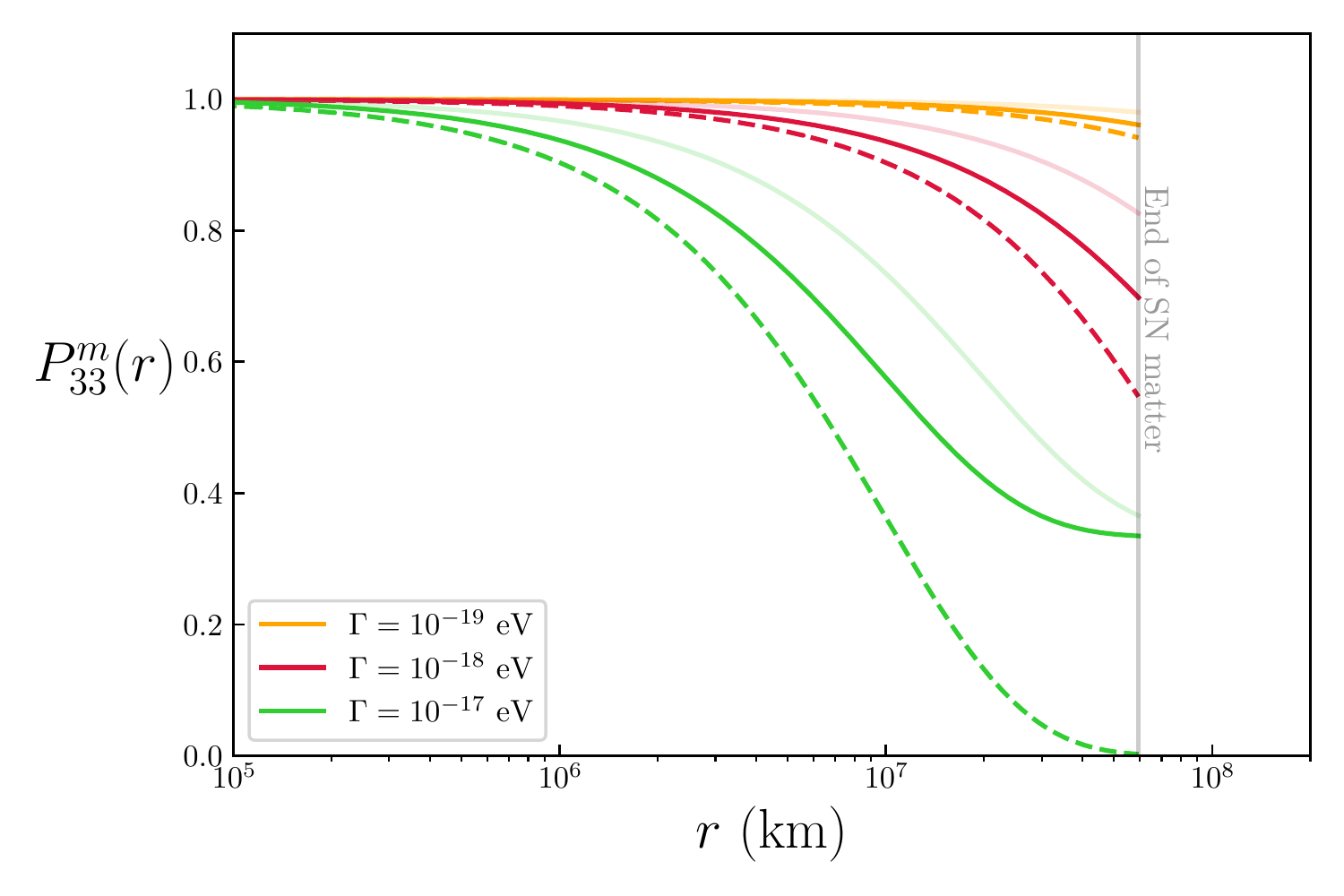}
    \caption{Solution for a survival probability of mass state 3 along a SN radius for the MSC$^{\epsilon}$ (solid opaque line) and neutrino loss (dashed). The transparent line shows the same probability but in vacuum. More details about these models are in the text. As it will be clear in our results, even with enhancement of the conversion in matter, values of $\Gamma \sim 10^{-19}$ eV are far higher than the sensitivity of a future SN detection compared to coherence length in vacuum used in the MSC$^\epsilon$ model.}
    \label{fig:P33sn}
\end{figure}

In Fig.~\ref{fig:P33sn} we compare the $P_{33}^{m(\text{SN})}$ to the same probability in mass basis in vacuum, which is shown as an enhancement of the deviation from the standard expectation of $P_{33}^{m(\text{SN})} = 1$. In Fig.~\ref{fig:Pii-conserved-E} we show the numerical probabilities of MSC$^{\slashed{\epsilon}}$ for the mass state in matter solved as described above.

\pagebreak
\FloatBarrier
\section{Tables with QD bounds}\label{appendix:b}

\begin{table}[h!]
\centering
\caption{Constrains for each detector for MSC$^\slashed{\epsilon}$ scenario with 90\%($2\sigma$) C.L. in units of $\Gamma \times 10^{-15}$ (eV). For $n \neq 0$ a representative energy of $E_0 = 10$~MeV was chosen and QD parameters are in eV scale. Values are corresponding to the simulated progenitor of 40~$M_\odot$.}
\label{tab:results-msc-noe}
\begin{tabular}{ll|lll|lll}
\hline
\multicolumn{2}{c}{} & \multicolumn{3}{c}{NH} & \multicolumn{3}{c}{IH} \\
\hline
Detector & SN distance & $n = 0$ & $n = 2$ & $n = 5/2$ & $n = 0$ & $n = 2$ & $n = 5/2$ \\
\hline
DUNE & 1 kpc & $0.89 (1.1) $ & $0.76 (0.89)$ & $0.65 (0.87)$ & $0.88 (1.0) $ & $2.5 (8.8) $ & $3.2 (15) $\\
 & 5 kpc & $5.4 (7.0) $ & $4.4 (5.9) $ & $6.3 (8.7) $ \\
 & 7 kpc & $8.3 (11) $ & $7.0 (9.4) $ & $11 (16) $ \\
 & 10 kpc & $14 (20) $ & $12 (17) $ & $22 (35) $ \\
HK & 1 kpc & $0.96 (1.1) $ & $3.7 (4.1) $ & $5.0 (5.8) $ & $0.93 (1.1) $ & $3.9 (4.3) $ & $5.3 (6.5) $ \\
 & 5 kpc & $4.3 (5.7) $ & $16 (21) $ & $33 (47) $ & $4.9 (6.6) $ & $18 (23) $ & $38 (49) $\\
 & 7 kpc & $7.1 (11) $ & $27 (38) $ & $53 (87) $ & $8.5 (13) $ & $28 (38) $ & $67 (99) $ \\
& 10 kpc & $16 (51) $ & $65 (120)$ & $150 (400)$ & $20 (36) $ & $52 (80) $ & $140 (240)$  \\
JUNO & 1 kpc & $4.2 (5.4) $ & $15 (19) $ & $30 (41) $ & $7.2 (8.9) $ & $38 (51) $ & $100 (180) $ \\
\hline
\end{tabular}
\end{table}

\begin{table}[h!]
\centering
\caption{Same as Table~\ref{tab:results-msc-noe} but for MSC$^\epsilon$ scenario with $2\sigma$($3\sigma$) C.L. in units of $\Gamma_8 \times 10^{-28}$ (eV). The representative energy of $E_0 = 10$~MeV was taken for $n \neq 0$ and QD parameters are in eV scale. Values are corresponding to the simulated progenitor of 27~$M_\odot$.}
\label{tab:results}
\begin{tabular}{ll|lll|lll}
\hline
\multicolumn{2}{c}{} & \multicolumn{3}{c}{NH} & \multicolumn{3}{c}{IH} \\
\hline
Detector & SN distance & $n = 0$ & $n = 2$ & $n = 5/2$ & $n = 0$ & $n = 2$ & $n = 5/2$ \\
\hline
DUNE & 1 kpc & $2.1 (3.3) $ & $0.43 (0.67)$ & $0.24 (0.37)$ & $490 (700) $ & $50 (180) $ & $33 (110) $\\
 & 5 kpc & $2.8 (5.2) $ & $0.58 (1.1) $ & $0.34 (0.75) $ \\
 & 7 kpc & $3.2 (7.1) $ & $0.71(1.9) $ & $0.46 (1.4) $ \\
 & 10 kpc & $4.2 (17) $ & $1.1 (7.0) $ & $0.80 (6.1) $ \\
HK & 1 kpc & $6.8 (11) $ & $0.81 (1.1) $ & $0.43 (0.58) $ & $9.2 (14) $ & $0.68 (1.0) $ & $0.36 (0.54) $ \\
 & 5 kpc & $9.6 (23) $ & $0.92 (1.9) $ & $0.48 (1.0) $ & $10 (18) $ & $0.80 (1.5) $ & $0.44 (0.82) $\\
 & 7 kpc & $13 $ & $1.2 $ & $0.61(2.6) $ & $11(25) $ & $0.94(2.5) $ & $0.51(1.4) $ \\
& 10 kpc & $42 $ & $2.7 $ & $1.2 $ & $13 $ & $1.4 $ & $4.9$  \\
JUNO & 1 kpc & $51 (100) $ & $6.4 (13) $ & $4.0 (7.8) $ & $47 (89) $ & $5.6 (11) $ & $3.5 (6.9) $ \\
\hline
\end{tabular}
\end{table}

\begin{table}[h!]
\centering
\caption{Same as Tables~\ref{tab:results-msc-noe} and \ref{tab:results} but for $\nu$-loss scenario, with $3\sigma$ bounds over $\gamma \times 10^{-29}$ (eV).}
\label{tab:results-loss}
\begin{tabular}{ll|lll}
\hline
\multicolumn{2}{c}{} & \multicolumn{3}{c}{IH} \\
\hline
Detector & SN distance & $n = 0$ & $n = 2$ & $n = 5/2$ \\
\hline
DUNE & 1 kpc & $500 $ & $4.6 $ & $2.1$ \\
 & 5 kpc & $100$ & $3.3$ & $1.6$ \\
 & 7 kpc & $74$ & $3.2$ & $1.5$ \\
 & 10 kpc & $52$ & $3.1$ & $1.5$ \\
HK & 1 kpc & $500$ & $2.6$ & $1.4$ \\
 & 5 kpc & $100$ & $2.3$ & $1.2$ \\
 & 7 kpc & $70$ & $2.2$ & $1.2$ \\
& 10 kpc & $49$ & $2.1$ & $1.2$ \\
JUNO & 1 kpc & $500$ & $150$ & $110$ \\
 & 5 kpc & $100$ & $32$ & $24$ \\
 & 7 kpc & $78$ & $24$ & $18$ \\
 & 10 kpc & $59$ & $19$ & $14$ \\
\hline
\end{tabular}
\end{table}



\end{document}